\documentclass[a4paper,fleqn,usenatbib]{mnras}

\usepackage{newtxtext,newtxmath}

\usepackage[T1]{fontenc}
\usepackage{ae,aecompl}

\usepackage[dvipdfmx]{graphicx}	

\usepackage{ulem}
\usepackage{color}
\newcommand{\fdir}{./}
\newcommand{\ncl}{N_{\rm cl}}
\newcommand{\mcl}{M_{\rm cl}}
\newcommand{\mbh}{m_{\rm bh}}
\newcommand{\mwd}{m_{\rm wd}}
\newcommand{\mcp}[1]{m_{#1}}
\newcommand{\ctde}{{\cal C}_{\rm tde}}
\newcommand{\ptde}{{\cal P}}
\newcommand{\ftde}{F_{\rm tde}}
\newcommand{\dcoal}{d_{\rm coal}}
\newcommand{\dtde}{d_{\rm tde}}
\newcommand{\rsch}{R_{\rm sch}}
\newcommand{\rstar}{R_{*}}
\newcommand{\mstar}{m_{*}}
\newcommand{\kd}{k_{\rm D}}
\newcommand{\nc}{n_{\rm c}}
\newcommand{\vc}{v_{\rm c}}
\newcommand{\rinfl}{r_{\rm infl}}
\newcommand{\dscl}{\hat{D}}
\newcommand{\msun}{M_\odot}
\newcommand{\rsun}{R_\odot}
\newcommand{\stde}{{\cal S}}
\newcommand{\rtde}{{\cal R}}
\newcommand{\rtded}{{\cal R}_{\rm d}}
\newcommand{\rtdedk}[1]{{\cal R}_{\rm d,#1}}
\newcommand{\rtdet}{{\cal R}_{\rm t}}
\newcommand{\rtdetk}[1]{{\cal R}_{\rm t,#1}}
\newcommand{\npcl}{{\cal N}}
\newcommand{\thubble}{T_{\rm Hubble}}
\newcommand{\sfr}{{\rm SFR}}
\newcommand{\mavi}{M_{{\rm av},i}}
\newcommand{\nsimi}{N_{{\rm sim},i}}
\newcommand{\msimi}{M_{{\rm sim},i}}
\newcommand{\yrgpc}{{\rm yr}^{-1}~{\rm Gpc}^{-3}}
\newcommand{\fbh}{f_{\rm BH}}
\newcommand{\fbhgc}{f_{\rm BH,GC}}
\newcommand{\ngc}{n_{\rm GC}}

\title[WD~TDEs in GCs and YMCs]{MOCCA-SURVEY Database -- I. Tidal
  disruption events of white dwarfs in globular clusters and young
  massive clusters}

\author[Ataru Tanikawa]{
Ataru Tanikawa$^{1}$\thanks{E-mail: tanikawa@ea.c.u-tokyo.ac.jp},
Mirek Giersz$^{2}$,
Manuel Arca Sedda$^{3}$
\\
$^{1}$Department of Earth Science and Astronomy, College of
  Arts and Sciences, The University of Tokyo, 3-8-1 Komaba, Meguro-ku,
  Tokyo 153-8902, Japan\\
$^{2}$Nicolaus Copernicus Astronomical Centre, Polish Academy of
  Sciences, ul. Bartycka 18, PL-00-716 Warsaw, Poland \\
$^{3}$Astronomisches Rechen-Institut, Zentrum f\"ur Astronomie der
  Universit\"at Heidelberg, M\"onchhoofstr. 12-14, D-69120 Heidelberg,
  Germany}

\date{Accepted XXX. Received YYY; in original form ZZZ}

\pubyear{2017}

\begin{document}
\label{firstpage}
\pagerange{\pageref{firstpage}--\pageref{lastpage}}
\maketitle

\begin{abstract}

We exploit more than 1000 star cluster Monte Carlo models from the
MOCCA-SURVEY Database I, to infer the local rate density of white
dwarf (WD) tidal disruption events (TDEs) in globular clusters (GCs)
and young massive clusters (YMCs). We suggest that the WD TDE rate for
GCs and YMCs in the local universe is $\sim 90$-$500~\yrgpc$, with 90
\% of WD TDEs occurring in GCs. The total WD TDE rate density is $\sim
9$-$50$ times larger than estimated before. Our results show that
thermonuclear explosions induced by WD TDEs can be observed at a rate
of $\sim 100$-$550~{\rm yr}^{-1}$ by the next generation optical
surveys, such as the Legacy Survey of Space \& Time by the Vera C.
Rubin Observatory. We also find that massive WDs are preferentially
disrupted due to mass segregation, and that 20 \% of exploding WDs
have $\gtrsim 1.0 \msun$ despite of the small population of such WDs.

\end{abstract}


\section{Introduction}
\label{sec:Introduction}

A star closely approaching a black hole (BH) can be torn apart by the
gravitational pull and undergo a so-called tidal disruption event
(TDE) \citep[for recent reviews,
  see][]{2015JHEAp...7..148K,2019GReGr..51...30S,2020arXiv200512528R}.
A subclass of TDEs involves the disruption of a white dwarf (WD), thus
being referred to as WD TDE \citep[see review
  by][]{2020SSRv..216...39M}.

WD TDEs can be a promising target of multi-messenger astronomy. As
electromagnetic (EM) sources, they have been suggested as the origin
of high-energy transients associated with gamma-ray emissions
\citep{2011ApJ...743..134K, 2014ApJ...781...13L, 2014ApJ...794....9M,
  2016ApJ...833..110I}. BHs accrete WD debris, and can accompany X-ray
flares \citep{2013ApJ...779...14J, 2019ApJ...871L..17S}, and
optical/ultraviolet transients \citep{2011ApJ...726...34C}. When a WD
deeply penetrates within the tidal radius, it can experience
thermonuclear explosion, and can be observed as a subtype of type Ia
supernovae (SNe Ia) \citep{1989A&A...209..103L, 2008CoPhC.179..184R,
  2009ApJ...695..404R, 2016ApJ...819....3M, 2017ApJ...839...81T,
  2018MNRAS.475L..67T, 2018ApJ...858...26T, 2018MNRAS.477.3449K,
  2020ApJ...890L..26K, 2018ApJ...865....3A, 2019ApJ...885..136A}, or
calcium-rich gap transients \citep{2015MNRAS.450.4198S,
  2018MNRAS.475L.111S}. \cite{2019MNRAS.487.2505K} have claimed a WD
TDE as the origin of a rapidly rising and luminous transient AT2018cow
\citep{2018ApJ...865L...3P}.  Aside from EM waves, WD TDEs should emit
gravitational waves (GWs) just before WDs are disrupted
\citep{2008MNRAS.391..718S}. Since WDs are much less compact than
neutron stars and BHs, the GW frequency does not match with
ground-based GW observatories, such as LIGO, Virgo, and KAGRA
\citep[][respectively]{2015CQGra..32g4001L, 2015CQGra..32b4001A,
  2019NatAs...3...35K}. LISA \citep{2017arXiv170200786A} may detect GW
emissions from WD TDEs, if the WD TDEs occur in the local group
\citep{2009ApJ...695..404R,
  2018ApJ...865....3A}. \cite{2020CQGra..37u5011A} have shown that GW
observatories with sensitivity at a decihertz frequency, such as ALIA
and DECIGO \citep[][respectively]{2019BAAS...51g.243M,
  2011CQGra..28i4011K}, can discover GWs of WD TDEs even in the
high-redshift universe. Several studies have suggested that WD TDEs
can be sources of ultra high-energy cosmic rays and neutrinos
\citep{2017PhRvD..96f3007Z, 2017PhRvD..96j3003A, 2018NatSR...810828B}.

WD TDEs represent a powerful way to probe the mass range of
intermediate-mass black holes (IMBHs), i.e. $10^2$ -- $10^4 \msun$
\citep{1989A&A...209..103L, 2009ApJ...705L.128R, 2018MNRAS.477.3449K}.
In fact, heavier IMBHs would directly swallow the WD, whilst lighter
BHs would most likely penetrate the WD and form a variant of a
Thorne-Zytkow object \citep[TZO][]{1975ApJ...199L..19T}. Even though
IMBHs should be a key to solve the formation of massive BHs (MBHs)
with $\gtrsim 10^5 \msun$ \citep[e.g.][]{1984ARA&A..22..471R,
  2001ApJ...562L..19E, 2012Sci...337..544V}, a small number of IMBHs
(or their promising candidates) have been discovered so far, such as
M82 X-1 \citep{2001ApJ...547L..25M}, HLX-1
\citep{2009Natur.460...73F}, free-floating IMBH candidates observed in
the Milky Way (MW) \citep{2014ApJ...783...62T, 2018ApJ...859...86T,
  2019ApJ...871L...1T, 2020ApJ...890..167T}, and the merger remnant of
a binary BH GW190521 \citep{2020PhRvL.125j1102A, 2020ApJ...900L..13A}.

Until today, no conclusive WD TDE has been reported, although there
are several candidates. This might owe to observational limits or more
fundamental reasons intrinsically connected with the co-evolution of
IMBHs and their stellar neighborhoods. Thus, it is crucial to
construct a theoretical framework to compare with future observation
surveys, like the one attainable with the Legacy Survey of Space \&
Time (LSST) by the Vera C. Rubin Observatory
\citep{2019ApJ...873..111I}. The observational features of a WD TDE
are expected to largely depend on the WD and BH properties. On the one
hand, accretion-driven flares are supposed to increase luminosity at
increasing the BH mass, assuming Eddington-limited accretion. On the
other hand, the WD composition, which determine the explosive
mechanisms and the observational features, varies with its mass, so
that from lighter to heavier WDs we have helium (He), carbon oxygen
(CO), and oxygen-neon (ONe) WDs.

WD TDEs can happen in two types of stellar systems harboring
IMBHs. The first type is represented by dwarf galaxies
\citep{2013ApJ...775..116R}. They are expected to contain IMBHs, since
they are on the extension line of the relation between MBHs and their
host galaxies \citep{1995ARA&A..33..581K, 1998AJ....115.2285M,
  2000ApJ...539L...9F, 2013ARA&A..51..511K}. The second type is
represented by star clusters, such as globular clusters
\citep[GCs][]{2012ApJ...758...28J} and young massive clusters
\citep[YMCs][]{2010ARA&A..48..431P}. IMBHs' seeds and themselves are
theoretically considered to form through repeated mergers of stars or
stellar-mass BHs
\citep[e.g.][]{2004Natur.428..724P,2015MNRAS.454.3150G}.

In this paper, we focus on WD TDEs in YMCs and GCs. Previous studies
investigated the development of WD TDEs in numerical simulations of
star clusters \citep{2004ApJ...613.1143B,
  2016ApJ...819...70M}. However, in the majority of these studies the
IMBH was placed in the cluster center ab-initio, thus avoiding the
complex phases affecting the IMBH seeding and growth. Here, we exploit
the MOCCA-SURVEY Database I, a suite of 2000 Monte Carlo simulations
representing star clusters with masses in the range $10^4-10^6
\msun$. These simulations couple dynamics and stellar evolution of
single and binary stars, thus enabling us to closely follow the
possible formation of an IMBH and development of a WD TDE. Moreover,
the large amount of available MOCCA models permit us to dissect the
properties of the BHs and WDs involved in a TDE. The MOCCA-SURVEY
Database I have been already used widely to study star clusters from
various points \citep{2016MNRAS.462.2950B, 2017MNRAS.464.4077B,
  2019MNRAS.483..315B, 2017MNRAS.464L..36A, 2017MNRAS.464.3090A,
  2018MNRAS.478.1844A, 2018ApJ...855..124S, 2018MNRAS.479.4652A,
  2019arXiv190500902A, 2018MNRAS.481.2168M, 2019arXiv190403591G,
  2019MNRAS.487.2412G, 2020MNRAS.498.4287H, 2021MNRAS.501.5212L}.

This paper is structured as follows. In section \ref{sec:Method}, we
introduce MOCCA-SURVEY Database I, and describe a method to calculate
the total event rate of WD TDEs and its derivatives. In sections
\ref{sec:Results} and \ref{sec:Comparison}, we show the results, and
compare the results with previous studies, respectively. In section
\ref{sec:Detectability}, we discuss about detectability of WD TDEs,
using our results. In section \ref{sec:Summary}, we summarize this
paper.

\section{Method}
\label{sec:Method}

We use MOCCA-SURVEY Database I in order to derive the event rate of WD
TDEs in GCs and YMCs. Before we introduce the database, we briefly
describe the MOCCA (MOnte Carlo Cluster simulAtor) code generating the
database. MOCCA clusters have been generated thanks to MOCCA, a Monte
Carlo code based on H\'enon's implementation
\citep{1971Ap&SS..14..151H} to follow the dynamical evolution of star
clusters, and includes improvements by \cite{1986AcA....36...19S} and
\cite{2008MNRAS.388..429G, 2013MNRAS.431.2184G}. It can deal with
single and binary star evolutions provided by
\cite{2000MNRAS.315..543H} and \cite{2002MNRAS.329..897H},
respectively. The MOCCA code contains the FEWBODY code
\citep{2004MNRAS.352....1F} as a module in order to solve close
encounters among single and binary stars by direct integration. The
MOCCA code models galactic tidal fields with reference to the theory
of potential escapers developed by \cite{2000MNRAS.318..753F}.

\begin{table*}
  \centering
  \caption{Initial conditions and setup of MOCCA-SURVEY Database I.}
  \label{tab:MoccaSurveyDatabaseI}
  \begin{tabular}{ccccccccccc}
    \hline
    $N/10^5$ & $R_{\rm t}$ [pc] & $R_{\rm t}/R_{\rm h}$ & $W_0$ & $Z/0.02$ & $f_{\rm b}$ & Natal kicks & $\ncl$ \\
    \hline
    $0.4$ & $30, 60, 120$ & $25,50$, Filling & $3.0,6.0,9.0$ & $0.05, 0.25, 1.0$ &
    $0.05, 0.1, 0.3, 0.95$ & Fallback & $243$ ($243$) \\
    $1.0$ & $30, 60, 120$ & $25,50$, Filling & $3.0,6.0,9.0$ & $0.05, 0.25, 1.0$ &
    $0.05, 0.1, 0.3, 0.95$ & Fallback & $268$ ($268$) \\
    $4.0$ & $30, 60, 120$ & $25,50$, Filling & $3.0,6.0,9.0$ & $0.05$ &
    $0.1, 0.3, 0.95$       & Fallback & $83$ ($64$) \\
    $7.0$ & $30, 60, 120$ & $25,50$, Filling & $3.0,6.0,9.0$ & $0.01, 0.05, 0.25, 0.3, 1.0$ &
    $0.1, 0.3, 0.95$       & Fallback & $325$ ($254$) \\
    $12$  & $30, 60, 120$ & $25,50$, Filling & $3.0,6.0,9.0$ & $0.05$ &
    $0.05, 0.1,  0.95$     & Fallback & $96$ ($71$) \\
    \hline
  \end{tabular}
\end{table*}

We present initial conditions and setup of the database in
Table~\ref{tab:MoccaSurveyDatabaseI}, where $N$ is the initial total
number of single and binary stars in a cluster, $R_{\rm t}$ and
$R_{\rm h}$ are the initial tidal and half-mass radii of a cluster,
$W_0$ is the initial concentration parameter of King's model
\citep{1966AJ.....71...64K} adopted for the initial phase space
distribution of each cluster, $Z$ is the metallicity of a cluster,
$f_{\rm b}$ is the initial binary fraction of a cluster, and $\ncl$ is
the number of clusters. A stellar initial mass function (IMF) in each
cluster is Kroupa's IMF \citep{2001MNRAS.322..231K} in which the
minimum and maximum stellar masses are $0.08 \msun$ and $100 \msun$,
respectively. We adopt the model of \cite{2002ApJ...572..407B} for
supernova model. Supernova natal kick velocities for neutron stars
(NSs) are drawn from a Maxwellian distribution with a dispersion of
265~km~s$^{-1}$ \citep{2005MNRAS.360..974H}. BH natal kicks are
modified by the mass fallback described by
\cite{2002ApJ...572..407B}. We select only models with BH natal kick
modified by the mass fallback, and thus our sample is limited to about
1000 MOCCA models, as summarized in Table 1. Each cluster is embedded
in an external tidal field by a point-mass galaxy with the flat
rotation velocity, $220$~km~s$^{-1}$.  According to the adopted galaxy
model, several MOCCA clusters can be dragged to the galactic center
via dynamical friction \citep{1975ApJ...196..407T,
  1993ApJ...415..616C, 2014ApJ...785...71G, 2014ApJ...785...51A}, a
feature that is not implemented in the MOCCA code. Therefore, we
exclude from our analysis all clusters with a dynamical friction
timescale <10 Gyr \citep{2019arXiv190500902A}.  We do not account for
these clusters to count the event rate of WD TDEs in the current
GCs. In Table \ref{tab:MoccaSurveyDatabaseI}, we indicate with $\ncl$
the number of clusters that survive to the dynamical friction drag.
\cite{2021MNRAS.501.5212L} have shown that these cluster models match
pretty well the MW GC system. As YMCs, we select clusters with $N \le
10^5$ and $Z=0.02$ from the MOCCA models. This is because YMCs have at
most $\sim 10^5 M_\odot$ and at least $\sim 0.02$ or $Z_\odot$
\citep{2010ARA&A..48..431P}. In the following, we will thus consider
MOCCA models as representative of the typical population of GCs and
YMCs.

\begin{figure}
  \includegraphics[width=\columnwidth]{\fdir/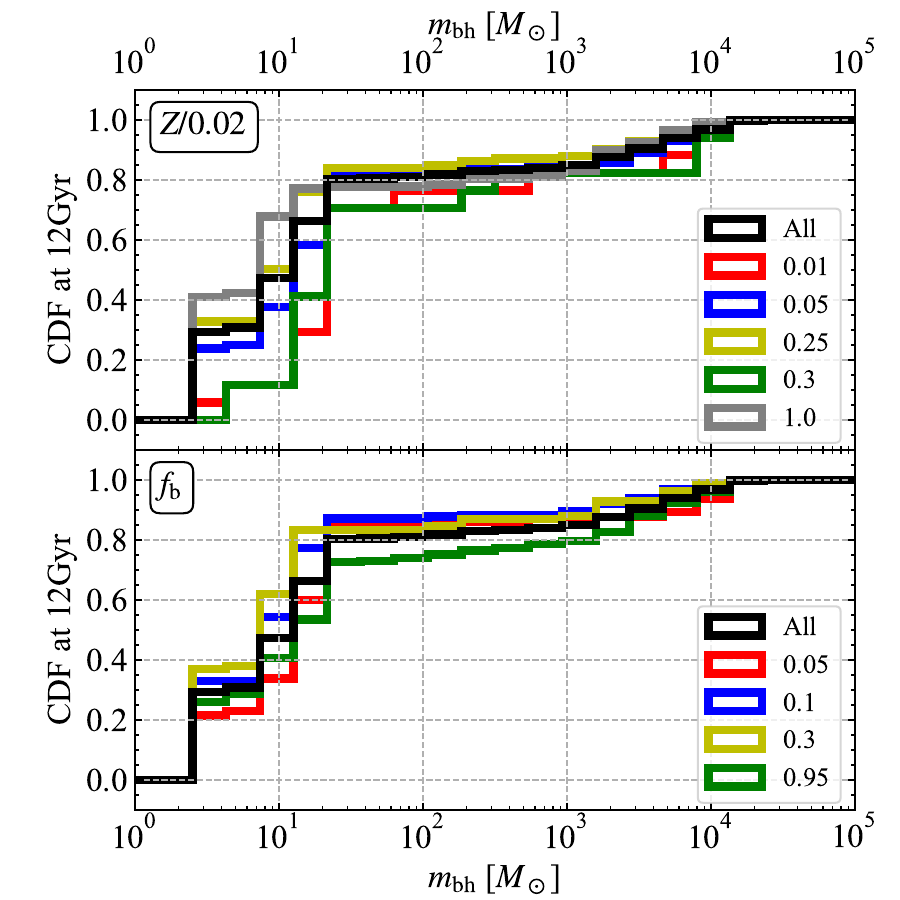}
  \caption{Cumulative distribution function (CDF) of the maximum BH
    mass in all the cluster models shown in Table
    \ref{tab:MoccaSurveyDatabaseI}. The ages of the clusters are 12
    Gyr old. 476 of 1015 clusters are disrupted or fall into galactic
    centers before this, and are not included in the CDFs. The black
    curves indicate the CDFs of all the cluster models, and other
    curves indicate the CDFs of cluster models with different
    metallicity (top) and initial binary fraction (bottom).}
  \label{fig:statisticsGc}
\end{figure}

Figure \ref{fig:statisticsGc} shows the cumulative distribution
function (CDF) of the maximum BH masses calculated after 12 Gyr in all
the examined models About 19, 16 and 4.8 \% of them contain BHs with
$>10^2$, $>10^3$, and $>10^4 \msun$ BHs, respectively. We set $10^2
\msun$ as the threshold to distinguish between stellar mass BHs and
IMBHs. According to this definition, the IMBH occupation fraction in
our models is $\fbhgc=0.19$. We note that this represents the most
optimistic case, as here we neglect the effect of GW recoil kick
imparted on BH merger remnants and the impact of the IMBH formation
scenario. We discuss a pessimistic scenario that accounts for these
features in Appendix \ref{sec:PessimisticCase}. The CDFs are not
sensitive to metallicity (Z) and initial binary fraction ($f_{\rm b}$)
as seen in the top and bottom panels of Figure \ref{fig:statisticsGc},
respectively. This indicates that the maximum BH mass function does
not much depend on our choice of the initial binary fractions and
metallicities of clusters.

We add more about $f_{\rm b}$. Several MOCCA models have an initial
binary fraction $f_{\rm b}$ = 0.95, which might appear unrealistic.
However, most of binary stars have wide separation, and are destroyed
by perturbation of other stars throughout cluster evolution. Then, the
binary fractions finally decrease to observational ones. Moreover,
such high $f_{\rm b}$ can rather explain observational properties of
cataclysmic variable stars \citep{2015MNRAS.446..226L,
  2016MNRAS.462.2950B, 2018MNRAS.474.3740B,
  2019MNRAS.483..315B}. Thus, $f_{\rm b}=0.95$ is not too large.

In the following, we consider only unbound WD-BH coalescence events,
labeling them as WD TDE candidates. We exclude from the analysis WD-BH
binary coalescence, because in this type of events the binary
undergoes stable mass transfer from the WD onto the BH over a
timescale much longer than WD TDE typical time.

\begin{figure}
  \includegraphics[width=\columnwidth]{\fdir/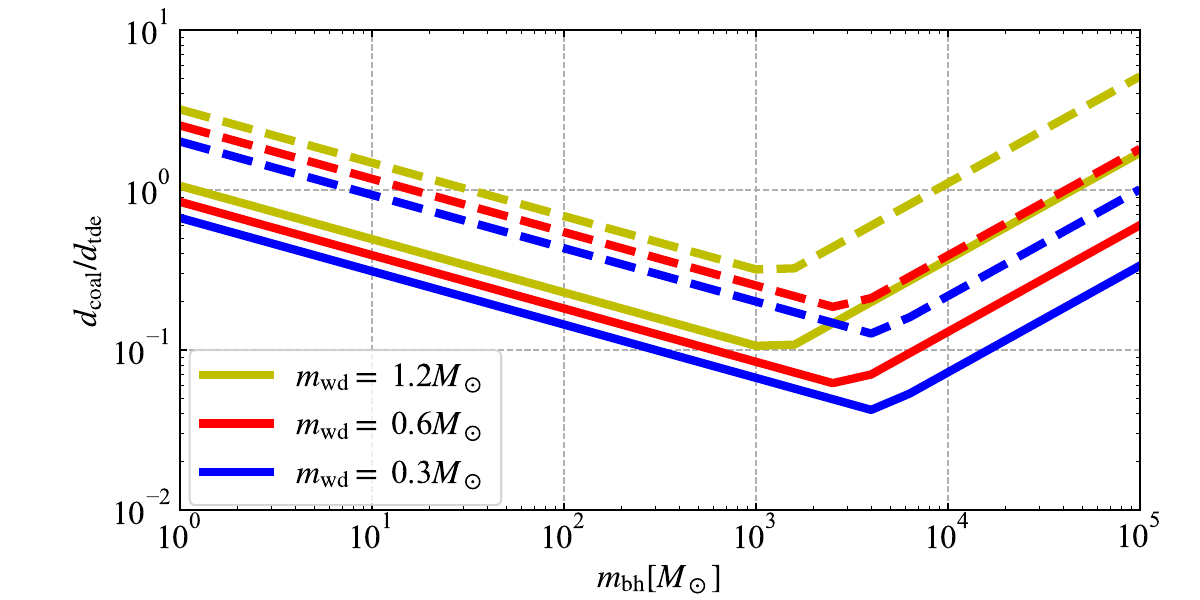}
  \caption{Ratio of a coalescence distance in the MOCCA models
    ($\dcoal$) to a tidal disruption distance ($\dtde$) as a function
    of BH mass ($\mbh$). Solid and dashed curves indicate the
    collision and few-body cases (see the main text),
    respectively. Colors indicate WD masses ($\mwd$).}
  \label{fig:radiusCollTde}
\end{figure}

In MOCCA models, a BH coalesces with a WD when their distance is less
than the sum of their radii, if they do not have companion stars
(hereafter, a collision case). Note that the BH radius is defined as
its Schwarzschild radius. A BH also coalesces with a WD when their
distance is less than 3 times the sum of their radii, if either of the
BH and WD have companion stars (hereafter, a few-body case).  Thus,
the MOCCA coalescence distance ($\dcoal$) can be written as
\begin{align}
  \dcoal &= \zeta \left( \rstar + \frac{2G\mbh}{c^2} \right),
\end{align}
and
\begin{align}
  \zeta &= \left\{
  \begin{array}{ll}
    1 & \mbox{(collision case)} \\
    3 & \mbox{(few-body case)}
  \end{array}
  \right..
\end{align}
where $G$ is the gravitational constant, $c$ is the light speed,
$\mbh$ is the BH mass, and $\rstar$ is the WD radius.  However, a WD
is tidally disrupted by a BH when their distance is less than a tidal
disruption distance, expressed as
\begin{align}
  \dtde = \eta \left( \frac{\mbh}{\mstar} \right)^{1/3}
  \rstar, \label{eq:TidalDisruptionDistance}
\end{align}
where $\eta$ is a numerical factor close to $1$, and $\mstar$ is the
WD mass. Figure \ref{fig:radiusCollTde} shows the ratio between the
MOCCA coalescence distance ($\dcoal$) and the tidal disruption
distance ($\dtde$) as a function of the BH mass. This ratio shows a
clear minimum at $\mbh \sim 10^3 \msun$. The ratio first decreases for
$\mbh \lesssim 10^3 \msun$, and next increases for $\mbh \gtrsim 10^3
\msun$. For the smaller-$\mbh$ range, $\dcoal$ depends on WD radii,
not BH radii. Thus, $\dcoal/\dtde$ decreases as $\dtde$ increases with
$\mbh$. For the larger-$\mbh$ range, $\dcoal$ is proportional to
$\mbh$. On the other hand, $\dtde$ is proportional to the cube root of
$\mbh$. Thus, $\dcoal/\dtde$ increases with $\mbh$. We see that the
$\dcoal/\dtde$ ratio is $<1$ in the whole $10^2-10^4 \msun$ range,
thus implying that $\dcoal$ is a likely too stringent condition to
identify WD TDE candidate and might lead to underestimate the actual
WD TDE rate.

It should be noted that even the case in which $\dtde$ equals $\dcoal$
would not imply necessarily a WD TDE. In fact, if the orbital
pericenter is smaller than the BH radius, the WD undergoes a direct
plunge without EM emission, whilst if the orbital pericenter is
smaller than the WD radius the BH will enter the WD structure and
settles in its center, forming a TZO \citep{1975ApJ...199L..19T},
whose EM emission is expected to significantly differ from a WD
TDE. We thus exclude direct plunges and TZO objects from our analysis.

Based on the aforementioned issue, we derive the number of WD TDEs in
two ways. In one way, we simply count the number of WD-BH coalescence
returned by MOCCA. In other words, the number of WD TDEs is the number
of WD-BH coalescence. In the other way, we multiply include in the
calculation a corrective factor defined as $(\dtde/\dcoal)\ftde$ (see
below). Here, $(\dtde/\dcoal)$ serves to take into account the fact
that the cross section of an interaction scales with the distance
between the two objects. The $\ftde$ factor, instead, excludes
plunging events and TZO objects, and it is defined as:
\begin{align}
  \ftde = \max \left[ 1 - \frac{\max \left( \rstar, \rsch
      \right)}{\dtde}, 0 \right], \label{eq:FractionOfTde}
\end{align}
where $\rsch$ is the Schwarzschild radius of the BH. The outer $\max$
of Eq. (\ref{eq:FractionOfTde}) indicates that a WD TDE cannot happen
if either of the WD and BH radii is larger than the tidal disruption
distance. The inner $\max$ of Eq. (\ref{eq:FractionOfTde}) decides
which object plunges into which object. The two ways give the lower
and upper bounds of the WD TDE rate with uncertainty from the
disagreement between the tidal disruption and coalescence
distances. We define the corrective factor, such that
\begin{align}
  \ctde = \left\{
  \begin{array}{ll}
    (\dtde/\dcoal)\ftde & \mbox{(case I)} \\
    1                   & \mbox{(case II)}
  \end{array}
  \right. \label{eq:CorrectionFactor}
\end{align}
The two corrective factors are representatives of two extreme
cases. The true corrective factor will be in between the two factors.
In the following, we focus on the case I, whilst we discuss the
outcomes of the case II in Appendix \ref{sec:Non-correctedWDTDERate}.

Once WD-BH coalescence events are retrieved from the MOCCA database,
we derive the WD TDE rate density ($\rtde$), i.e. the number of WD
TDEs per unit time and volume. Here, we describe a method to obtain
the WD TDE rate density differentiated by a set of WD TDE properties
$\ptde$ (hereafter, the differential rate density). Note that $\ptde$
represents several quantities, such as BH mass ($\mbh$) and WD mass
($\mwd$). The method is based on \cite{2004A&A...415..407B} and
\cite{2017MNRAS.464L..36A}. The differential rate density can be
calculated as
\begin{align}
  \frac{d\rtde}{d\ptde} = \sum_{i={\rm GC,YMC}} \int_0^{\thubble} dt
  \frac{\sfr_i(z(t))}{\mavi} \frac{d\npcl_i}{dt
    d\ptde}. \label{eq:DifferentialRateDensity}
\end{align}
As seen in Eq.~(\ref{eq:DifferentialRateDensity}), we calculate WD~TDE
rates in GCs and YMCs separately.  We define $\sfr_i(z(t))$ as the
cluster formation rate density in mass at the lookback time $t$, and
$\mavi$ as the average cluster mass.  We represent $\npcl_i$ as the
total number of WD~TDEs in a cluster. Thus, differentiating $\npcl_i$
by the cluster age $t$, we can get $d\npcl_i/dt$ as the WD~TDE rate at
the cluster age $t$, and $d\npcl_i/dt d\ptde$ as the WD~TDE rate per a
property at the cluster age $t$. Solving
Eq. (\ref{eq:DifferentialRateDensity}) thus returns the WD TDE rate
density for a given property (e.g. BH or WD mass, see below).

We discretize Eq.~(\ref{eq:DifferentialRateDensity}) in such a way
that the differential number of WD TDEs is given by:
\begin{align}
  \frac{d\npcl_i}{dt d\ptde} = \frac{1}{\nsimi} \sum_{j=1}^{N_i}
  \delta(t-t^j) \delta(\ptde-\ptde^j) \left( \prod_k \ctde^{jk}
  \right), \label{eq:DiscretizedRateDensityPerCluster}
\end{align}
where $\nsimi$ is the number of all the simulations, $N_i$ is the
total number of WD-BH coalescences in all the simulations, and
$\delta$ functions have usual meanings. The superscript $j$ indicates
quantities of $j$-th WD-BH coalescence in all the simulations. The
product of corrective factors considers that a WD TDE can accompany
another TDEs. Note that we can get the total number of WD-BH
coalescence in a cluster ($\npcl_i$), if we exclude the $\delta$
functions and the product of corrective factors. Finally, we can
obtain a discretized rate density in the current universe,
substituting Eq.~(\ref{eq:DiscretizedRateDensityPerCluster}) into
Eq.~(\ref{eq:DifferentialRateDensity}), such that
\begin{align}
  \frac{d\rtde}{d\ptde} = \sum_{i={\rm GC,YMC}} \left[ \sum_j^{N_i}
    \frac{\sfr_i(z(t^j))}{\msimi} \delta(\ptde-\ptde^j) \left( \prod_k
    \ctde^{jk} \right) \right] \label{eq:DiscretizedRateDensity},
\end{align}
where $\msimi(=\mavi \nsimi)$ is the total mass of clusters in all the
simulations. We give examples of use for
Eq. (\ref{eq:DiscretizedRateDensity}). We can get the differential
rate density of simple WD TDEs as
\begin{align}
  &\frac{d\rtde}{d\mbh d \mwd} \nonumber \\
  &= \sum_{i={\rm GC,YMC}} \left[ \sum_j^{N_i}
    \frac{\sfr_i(z(t^j))}{\msimi} \delta(\mbh-\mbh^j)
    \delta(\mwd-\mwd^j) \ \ctde^{j} \right].
\end{align}
On the other hand, we can express the differential rate density of WD
TDEs with another TDE as
\begin{align}
  \frac{d\rtde}{d\mbh d\mwd d\mstar} &= \sum_{i={\rm GC,YMC}} \Biggl[
    \sum_j^{N_i} \frac{\sfr_i(z(t^j))}{\msimi} \delta(\mbh-\mbh^j)
    \nonumber \\
    &\times \delta(\mwd-\mwd^j)
    \delta(\mstar-\mstar^j) \ctde^{\rm wd} \ctde^{*} \Biggr],
\end{align}
where $\mstar$ is the mass of another star disrupted simultaneously
with the WD.

We also define a volume-limited WD TDE rate as
\begin{align}
  \stde = \int_0^{z_{\max}} \rtde(z) \frac{dV(z)}{dz}
  dz, \label{eq:VolumeLimitedRate}
\end{align}
where $\rtde(z)$ and $V(z)$ are the WD TDE rate density and comoving
volume at a redshift of $z$, respectively. Note that we can derive
$\rtde(z)$ or $d\rtde(z)/d\ptde$ when we replace $\thubble$ with
$\thubble - t_{\rm lb}(z)$ in Eq. (\ref{eq:DifferentialRateDensity}),
where $t_{\rm lb}(z)$ is a lookback time at a redshift of $z$. For
confirmation, the unit of $\stde$ is the number of WD TDEs per unit
time.

We adopt the GC formation rate, $\sfr_{\rm GC}$, estimated by
\cite{2013MNRAS.432.3250K}, and simplify their GC formation rate as
\begin{align}
  \sfr_{\rm GC}(z) = \left\{
  \begin{array}{ll}
    0.003 \msun~{\rm yr}^{-1}~{\rm Mpc}^{-3}  & (2 \le z \le 8) \\
    0                                         & ({\rm otherwise})
  \end{array}
  \right..
\end{align}
We set the YMC formation rate, $\sfr_{\rm YMC}$, as
\begin{align}
  \sfr_{\rm YMC}(z) = 0.016 f_{\rm YMC} \psi(z) \msun~{\rm
    yr}^{-1}~{\rm Mpc}^{-3}, \label{eq:YMCFormationRate}
\end{align}
where
\begin{align}
  \psi(z) = \frac{(1+z)^{2.6}}{1+\left[(1+z)/3.2\right]^{6.2}}.
\end{align}
We assume the current star formation rate as $0.016 \msun~{\rm
  yr}^{-1}~{\rm Mpc}^{-3}$ \citep[e.g.][]{2013ApJ...770...57B}, and
set the fraction of YMCs to $f_{\rm YMC} = 0.03$. The redshift
dependence is the same as the star formation rate derived by
\cite{2017ApJ...840...39M}. We estimate the fraction of YMCs by
restricting ourselves to those with an age $<1$ Gyr, under the
assumption that older ones will have evaporated
\citep{2019ARA&A..57..227K}. We ignore WD TDEs at a cluster age of
more than $1$ Gyr when calculating the WD TDE rate density in YMCs.

\begin{table}
  \centering
  \caption{Properties of GCs and YMCs in the database}
  \label{tab:PropertiesOfModels}
  \begin{tabular}{cccc}
    \hline
    Clusters & $\nsimi$ & $\msimi$ [$\msun$] & $M_{\rm av}$ [$\msun$] \\
    \hline
    GC  & $1015$ & $3.0 \times 10^8 \msun$ & $2.8 \times 10^5 \msun$ \\
    YMC &  $174$ & $8.8 \times 10^6 \msun$ & $5.1 \times 10^4 \msun$ \\
    \hline
  \end{tabular}
\end{table}

For $\npcl_{\rm GC}$, we adopt clusters surviving from dynamical
friction. For $\npcl_{\rm YMC}$, we use clusters with $N \le 10^5$ and
$Z=0.02$ as described above. We also show $\nsimi$, $\msimi$, and
$\mavi$ of GCs and YMCs in Table~\ref{tab:PropertiesOfModels}. We
obtain the local number density of GCs as $29.1$~Mpc$^{-3}$, combining
$\sfr_{\rm GC}$ with $M_{\rm av,GC}$, if all the GCs survive. In fact,
$\sim 53$ \% of cluster models selected for GCs survive until 12 Gyr
old. This means that the current GC number density ($\ngc$) is
$15.4$~Mpc$^{-3}$. Our value of $\ngc$ is considerably larger than the
estimate provided by \cite{2000ApJ...528L..17P} and
\cite{2020SSRv..216...39M} or the simple estimate that can be inferred
from MW GCs \citep[e.g.][]{2016ApJ...824L...8R}. Since $\ngc$
represents a scale factor in our calculations, we explicitly leave
this dependence in all our calculations, adopting a fiducial value of
$\ngc = 15.4$~Mpc$^{-3}$. Similarly, we infer that the number density
of YMCs is $9.45$ Mpc$^{-3}$.

\begin{table}
  \centering
  \caption{Models to calculate WD TDE rate densities}
  \label{tab:ModelName}
  \begin{tabular}{lll}
    \hline
    Name & IMBH formation & Eq. (\ref{eq:CorrectionFactor}) \\
    \hline
    Model A & optimistic case  & case I \\
    Model B & optimistic case  & case II \\
    Model C & pessimistic case & case I \\
    \hline
  \end{tabular}
\end{table}

Finally, we summarize three models (hereafter called models A, B, and
C) to calculate the WD TDE rate density (see also Table
\ref{tab:ModelName}). In model A, we consider the optimistic case for
the IMBH formation ($\fbhgc=0.19$ as seen in Figure
\ref{fig:statisticsGc}), and adopt the case I for the corrective
factor $\ctde$ in Eq. (\ref{eq:CorrectionFactor}). We focus on this
results in section \ref{sec:Results}. In model B, we consider the
optimistic case, the same as model A, however adopt the case II for
the corrective factor $\ctde$ in Eq. (\ref{eq:CorrectionFactor}). We
describe the results of model B in Appendix
\ref{sec:Non-correctedWDTDERate}. In model C, we consider the
pessimistic case for the IMBH formation, and adopt the case I for the
corrective factor $\ctde$ in Eq. (\ref{eq:CorrectionFactor}). This
results are written in Appendix \ref{sec:PessimisticCase}.

\section{Results}
\label{sec:Results}

We obtain the total rate density of WD TDEs ($\rtde$) as $4.9 \times
10^2 \yrgpc$ in model A, and $1.1 \times 10^2 \yrgpc$ in model B (see
Appendix \ref{sec:Non-correctedWDTDERate}). The former is 5 times as
large as the latter. This is because WD TDEs are dominated by WD TDEs
of single CO and ONe WDs and single BHs with $\mbh \sim 10^3$ -- $10^4
\msun$ as shown later, and their $\dcoal/\dtde$ is $\sim 0.2$ as seen
in Figure \ref{fig:radiusCollTde}. The rate density is $8.9 \times
10^1 \yrgpc$ in model C (see Appendix \ref{sec:PessimisticCase}),
which is also about 5 times less than the rate density for model A.
Hereafter, we focus on the rate densities and their derivatives
obtained in model A. We should bear in mind that they can be
overestimated by at most 5 times.

\begin{figure*}
  \includegraphics[width=1.5\columnwidth]{\fdir/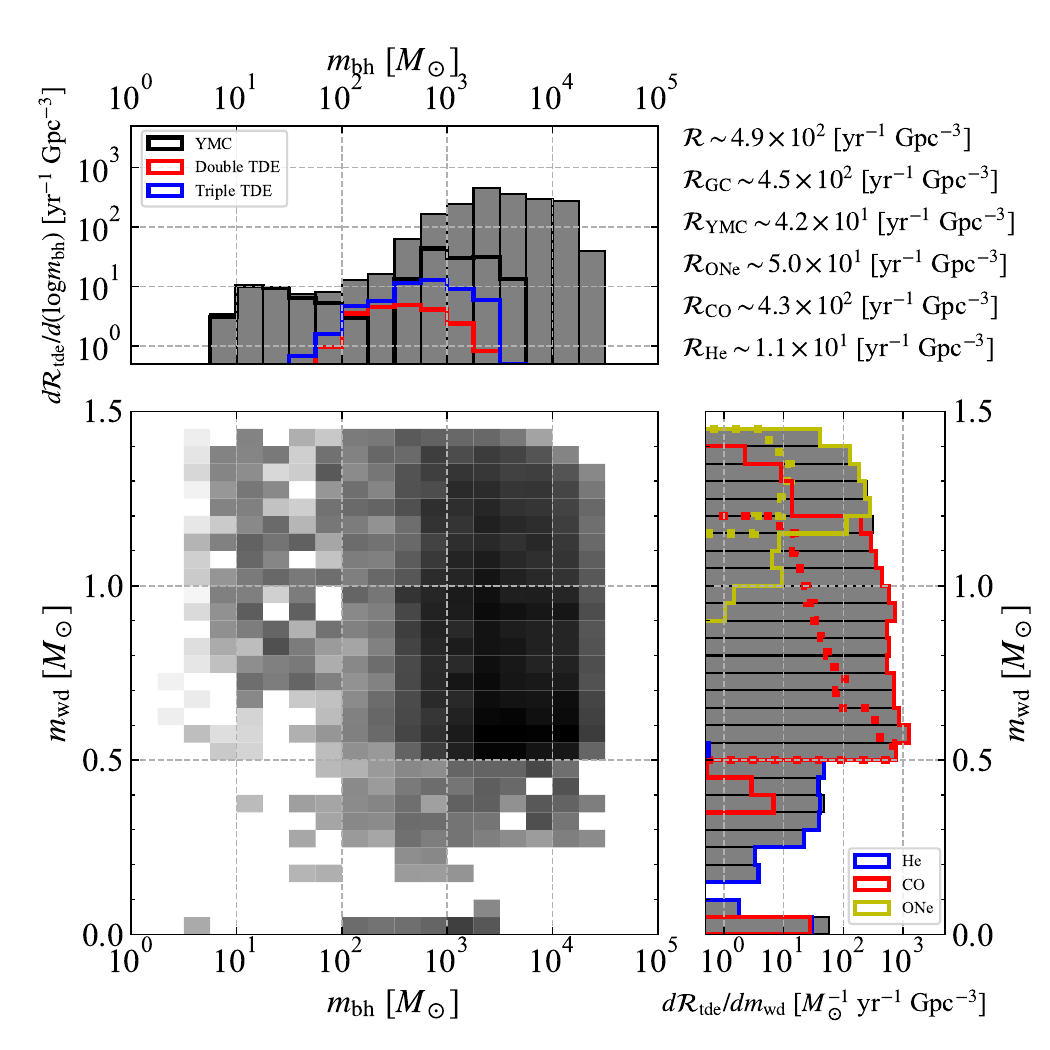}
  \caption{WD TDE rate density in the current GCs and YMCs. On the
    top-right conner, we show the rate densities of all WD TDEs
    (${\cal R}$), WD TDEs in GCs (${\cal R}_{\rm GC}$), WD TDEs in
    YMCs (${\cal R}_{\rm YMC}$), ONe WD TDEs (${\cal R}_{\rm ONe}$),
    CO WD TDEs (${\cal R}_{\rm CO}$), and He WD TDEs (${\cal R}_{\rm
      He}$). We call double and triple TDEs ``multiple TDEs'' all
    together.  A multiple TDE is defined as a phenomenon where a WD
    TDE happens along with another TDEs in a binary-single or
    binary-binary encounter. Another TDEs do not have to be WD
    TDEs. Multiple TDEs can contain two or three WD TDEs. We count
    them as twice, and three times WD TDEs. Bottom left: Relative
    abundance of the rate density per $\mwd$ per dex of $\mbh$ in a
    logarithmic color scale. Top left: The rate density per dex of
    $\mbh$. Gray-filled histograms indicate all the WD TDEs, and
    black-open histograms WD TDEs only in YMCs. Red- and blue-open
    histograms show WD TDEs in double and triple TDEs,
    respectively. Bottom right: The rate density per
    $\mwd$. Gray-filled histograms are the same as in the top left
    panel. Yellow-, red-, and blue-open histograms indicate ONe WD, CO
    WD, and He WD TDEs, respectively. Yellow- and red-dotted curves
    show the abundance of ONe WDs and CO WDs, respectively, yielded
    through single star evolution. We obtain this abundance as
    follows. We prepare $Z=0.02$ single stars with the Kroupa's IMF,
    and calculate their $15$ Gyr evolutions by the SSE code
    \citep{2000MNRAS.315..543H}. The abundance of $0.5-0.55 \msun$ CO
    WDs is normalized to the rate density per $\mbh$ at $0.5-0.55
    \msun$. He WDs are not formed within $15$ Gyr through single star
    evolution. }
  \label{fig:tdeTotal}
\end{figure*}

Figure \ref{fig:tdeTotal} shows the WD TDE rate density and its
breakdowns. As seen in the top left panel, a majority of WD TDEs
happen in BHs with $\mbh \gtrsim 10^3 \msun$. The fraction of WD TDEs
with $\mbh \lesssim 10^3 \msun$ is less than 10 \%. This is not
because there are more many clusters with $\gtrsim 10^3 \msun$ BHs
than with $\lesssim 10^3 \msun$ BHs, but because more massive BHs
yield WD TDEs much more frequently. This will be seen in
Eq. (\ref{eq:WdTdeRate2}) of section \ref{sec:Comparison}. The WD TDE
rate decreases for $\mbh \gtrsim 10^4 \msun$ in two steps. First, WDs
tend to be just swallowed by BHs without tidally disrupted. Second,
clusters cannot form BHs with more than several $10^4 \msun$.

As seen on the top right corner, more than 90 \% of WD TDEs happen in
GCs, and the rest in YMCs. This is not only because the local number
density of GCs is larger than that of YMCs, but also because GCs
harbor much more massive BHs than YMCs. GCs can have $\gtrsim 10^4
\msun$ BHs, while YMCs can have $\lesssim 10^4 \msun$ BHs (see the top
left panel of Figure \ref{fig:tdeTotal}). Since we assume that YMCs
have smaller masses and shorter lifetimes than GCs, YMCs cannot yield
$\gtrsim 10^4 \msun$ BHs. Interestingly, YMCs dominate WD TDEs with
$\lesssim 10^2 \msun$ BHs, although such WD TDEs are minor on the
whole of WD TDEs.

We can see the WD mass distribution from the bottom right panel of
Figure \ref{fig:tdeTotal}. The WD TDE rate suddenly increases above
$\mwd \sim 0.5 \msun$. $\gtrsim 0.5 \msun$ WDs (CO and ONe WDs) can be
formed through single star evolution within the Hubble time. On the
other hand, $\lesssim 0.5 \msun$ WDs (He WDs) cannot be formed through
single star evolution within the Hubble time, and can be formed only
through binary star evolution. For $\mwd \gtrsim 0.5 \msun$, the WD
TDE rate gradually decreases with $\mwd$ increasing for the following
two reasons. First, the Kroupa's IMF yields a smaller number of more
massive WDs. Second, more massive WDs can be more easily swallowed by
BHs without tidally disrupted. This can be clearly seen for $\mwd
\gtrsim 1.2 \msun$.

The abundance of He WD, CO WD, and ONe WD TDEs is $\sim 2.2$, $88$,
and $10$ \% of the total WD TDEs, respectively. In order to compare
this abundance with the abundance of WDs yielded through single star
evolution, we prepare $Z=0.02$ single stars with the Kroupa's IMF, and
calculate their $15$~Gyr evolutions by the SSE code
\citep{2000MNRAS.315..543H}. Then, the abundance of He, CO, and ONe
WDs is $0$, $97.5$, and $2.5$ \%, respectively. The former WD
abundance is quite different from the latter WD abundance. Single star
evolution cannot form He WDs for $15$ Gyr, while binary star evolution
forms He WDs, and yields He WD TDEs. The abundance of ONe WD TDEs is
$\sim 4$ times larger than the abundance of ONe WDs from single
stars. Moreover, the abundance of $\sim 1 \msun$ CO WD TDEs is $\sim
10$ times larger than the abundance of CO WDs from single stars,
compared between the solid and dashed red curves in the bottom right
panel of Figure \ref{fig:tdeTotal}. WD TDEs prefer to more massive
WDs, except that binary star evolution yield He WD TDEs.

\begin{figure}
  \includegraphics[width=\columnwidth]{\fdir/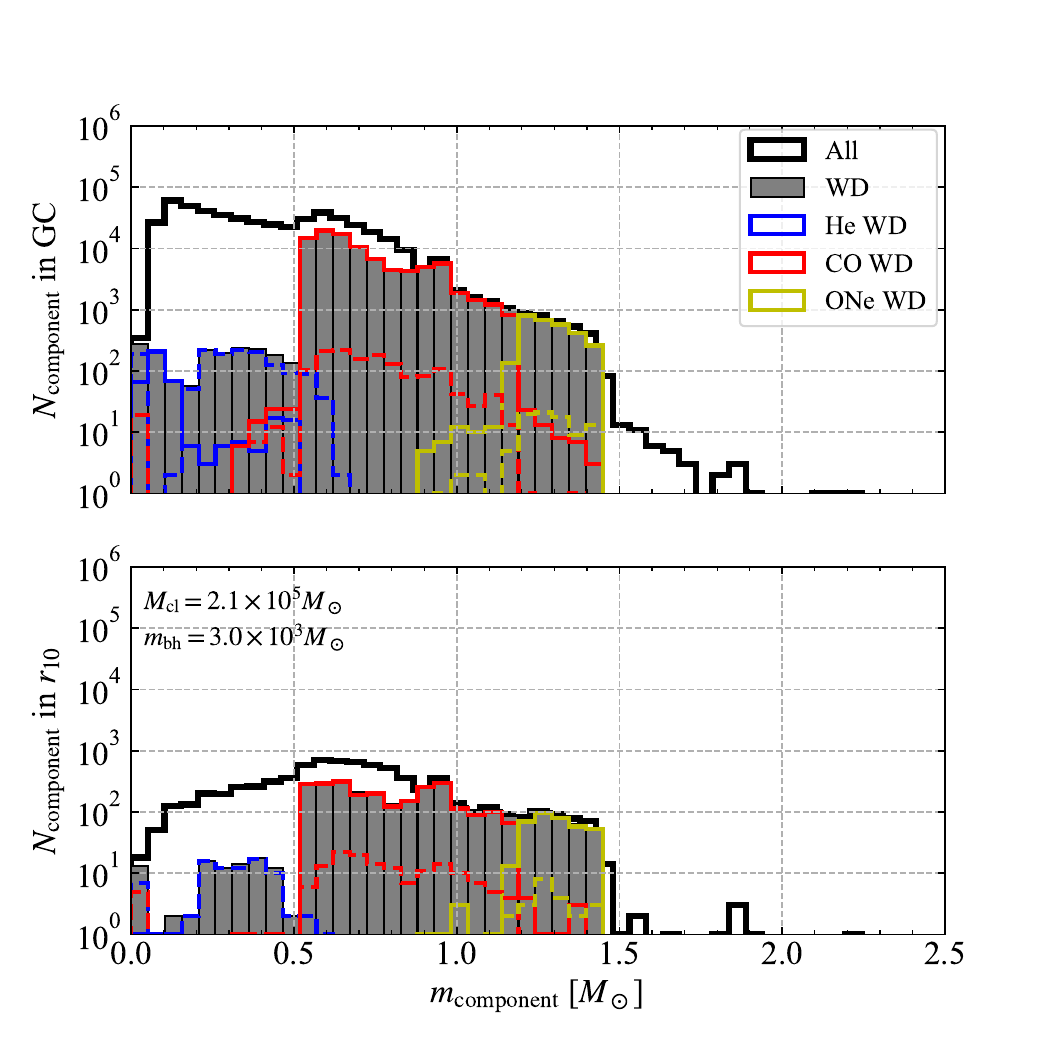}
  \caption{Number of stars in a cluster (top) and in its $10$ \%
    Lagrangian radius (bottom) as a function of stellar masses. The
    age of the cluster is $12$ Gyr, and the cluster has $N=7 \times
    10^5$, $R_{\rm t}=30$~pc, $R_{\rm t}/R_{\rm h}=50$, $W_0=6$, and
    $Z=0.001$ at the initial time.  We exclude the most massive BH to
    obtain the $10$ \% Lagrangian radius. We count a binary star as
    two stars. Solid black curves indicate all the stars. Filled
    histograms show all the WDs. Solid blue, red, and yellow curves
    indicate He, CO, and ONe WDs. Dashed blue, red, and yellow curves
    indicate He, CO, and ONe WDs in binaries.}
  \label{fig:massWdAsComponent}
\end{figure}

The reason why WD TDEs prefer to more massive WDs is mass segregation
by which more massive objects fall into the GC center more
preferentially. Figure \ref{fig:massWdAsComponent} shows the number of
stars in a GC and in its inner region as a function of stellar masses.
The number ratio of ONe WDs with $\sim 1.4 \msun$ to CO~WDs with $\sim
0.5 \msun$ is $1/30$ for the whole GC, and $1/3$ for its inner region.
Thus, more massive WDs experience WD TDEs more easily.

He WDs also sink into the inner region preferentially despite of their
small masses. The ratio of He WDs with $\sim 0.4 \msun$ to CO WDs with
$\sim 0.5 \msun$ is $1/100$ for the whole GC, and $1/30$ for the inner
region. This is because He WDs are likely to be binary members owing
to their formation mechanism. The total masses of these He~WDs and
their companions are larger than the surrounding stars, and these He
WDs sink into the inner region with their companions. This may be the
reason why He WD TDEs still happen frequently despite of the small
masses of He WDs.

$0$-$0.05 \msun$ CO WDs and He WDs equally contribute to the WD TDE
rate. Thus, the red curve overlaps on the blue curve in the smallest
histogram bin in the lower right panel of Figure
\ref{fig:tdeTotal}. Such low-mass CO WDs are formed as follows. A
binary star evolves to a close binary star with two CO WDs. The
separation of the close binary star shrinks due to gravitational wave
radiation, and the lighter CO WD fills its Roche lobe at some
point. When the mass ratio of the lighter CO WD to the heavier CO WD
is smaller than $\sim 0.6$, the mass transfer from the lighter CO WD
to the heavier CO WD is stable. Thus, the lighter CO WD gradually
loses its mass down to $\lesssim 0.05 \msun$ without the merger of the
two CO WDs. The tidally disrupted CO WDs have a mass gap between
$0.05$ and $0.3 \msun$. This is because the mass-loss timescale from
$\sim 0.3 \msun$ to $\sim 0.05 \msun$ is much shorter than that from
$\sim 0.05 \msun$ to $\sim 0 \msun$.

The top left panel of Figure \ref{fig:tdeTotal} also shows the rate
density of multiple TDEs (defined in the caption of Figure
\ref{fig:tdeTotal}). In multiple TDEs, a series of TDEs should happen
at time intervals of an order of the maximum periods of binaries
involving these encounters\footnote{This is just a rough estimate. If
  multiple TDEs occur in temporary (but long-lived) multiple stellar
  systems (so-called resonant encounters), the intervals of TDEs can
  be much more than the maximum periods of binaries.}.  We find that
$80$~\% of these binaries have periods of less than $100$ days. Thus,
multiple TDEs can be observed as single (but resolvable) transient
events at the same time and location.

\begin{figure}
  \includegraphics[width=\columnwidth]{\fdir/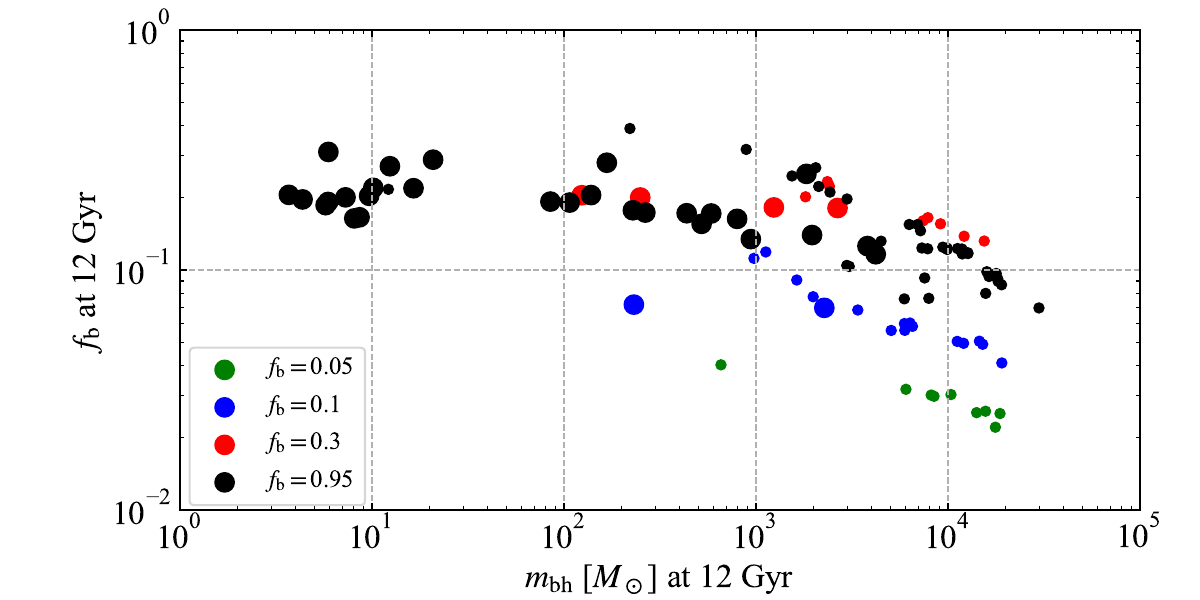}
  \caption{Binary fraction in the $10$ \% Lagrangian radius of each GC
    at $12$ Gyr as a function of the most massive BH mass. We exclude
    the most massive BH to obtain the $10$ \% Lagrangian radius. We do
    not plot a GC if it has a BH whose mass is half of the GC
    mass. Color coding indicates the binary fraction of each GC at the
    initial time. Large dots indicate GCs with multiple TDEs including
    at least one WD TDE, and small dots indicate GCs with WD TDEs but
    without multiple TDEs.}
  \label{fig:binaryFraction}
\end{figure}

Multiple TDEs are most likely to occur with $\sim 10^3 \msun$ BHs,
despite that the WD TDE rate increase with $\mbh$ increasing (see the
top left panel of Figure \ref{fig:tdeTotal}). This can be interpreted
as follows. Figure \ref{fig:binaryFraction} shows that the binary
fractions in the inner regions of GCs become smaller with $\mbh$
increasing. Since multiple TDEs need binary stars, they are harder to
happen in GCs with larger $\mbh$. In the process of BH growth, the BHs
disrupt binary stars by unbinding them, or by swallowing either of
them.

\begin{figure}
  \includegraphics[width=\columnwidth]{\fdir/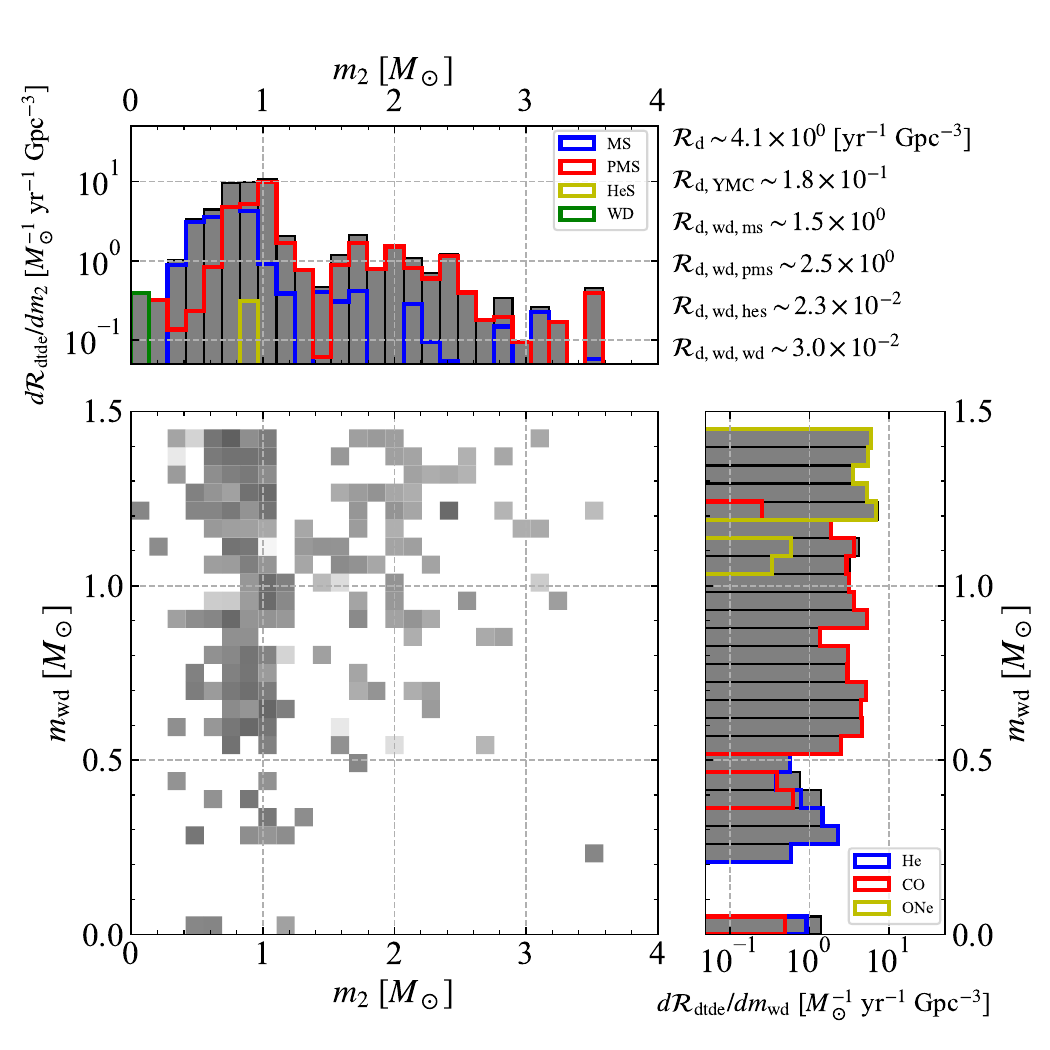}
  \caption{Double TDE rate density in the current GCs and YMCs. On the
    top-right corner, we show the rate densities of all double TDEs
    ($\rtded$), those in YMCs ($\rtdedk{YMC}$), those with WD and main
    sequence (MS) ($\rtdedk{wd,ms}$), those with WD and post MS (PMS)
    ($\rtdedk{wd,pms}$), those with WD and helium stars (HeS)
    ($\rtdedk{wd,hes}$), and those with two WDs
    ($\rtdedk{wd,wd}$). Bottom left: Relative abundance of the rate
    density per $\mwd$ per companion mass ($\mcp{2}$). Top left: The
    rate density per $\mcp{2}$. Gray-filled histograms indicate all
    the double TDEs. Colored-open histograms indicate stellar types of
    companion stars, such that blue-, red-, yellow-, and green-open
    histograms are MSs, PMSs, HeSs, and WDs, respectively. Bottom
    right: The rate density per $\mwd$. Gray-filled histograms
    indicate all the double TDEs. Colored-open histograms indicate WD
    types, such that blue-, red-, and yellow--open histograms are He,
    CO, and ONe WDs, respectively.}
  \label{fig:tdeDouble}
\end{figure}

Figure \ref{fig:tdeDouble} shows the double TDE rate density. The
total rate density is $\sim 4.1~\yrgpc$, i.e. only $\sim 0.84$ \% of
the total WD TDEs. The double TDEs in YMCs are only $4.3$ \% of the
total double TDEs. Most of the double TDEs consist of WDs and main
sequences (MSs), or of WDs and post MSs (PMSs). The reason for the
abundant MSs is that MSs have long lifetimes, and relatively large
radii $\sim 1 \rsun$. The reason for the abundance PMSs is that PMSs
have large radii $\gg 1 \rsun$, although their lifetimes are not long.

The rate density weakly depends on WD masses for $\mwd \gtrsim 0.5
\msun$, or for CO and ONe WDs. Mass segregation strongly affects the
WD mass function of double TDEs. The rate density with He WDs is
similar to those with CO and ONe WDs. This is because most of He WDs
have their companions due to their formation processes. The companion
mass distribution has a peak at $0.7-1.0 \msun$, which consists of MSs
and PMSs. The mass of $\sim 1.0 \msun$ corresponds to the turn-off
mass at $\sim 12$ Gyr. The number of $>1.0 \msun$ MS and PMS stars is
small, since these stars already evolve to stellar
remnants. Nevertheless, double TDEs can contain $> 1.0 \msun$ MSs and
PMSs. These stars experience stellar coalescence or mass transfer from
their companions before they are tidally disrupted.

\begin{figure}
  \includegraphics[width=\columnwidth]{\fdir/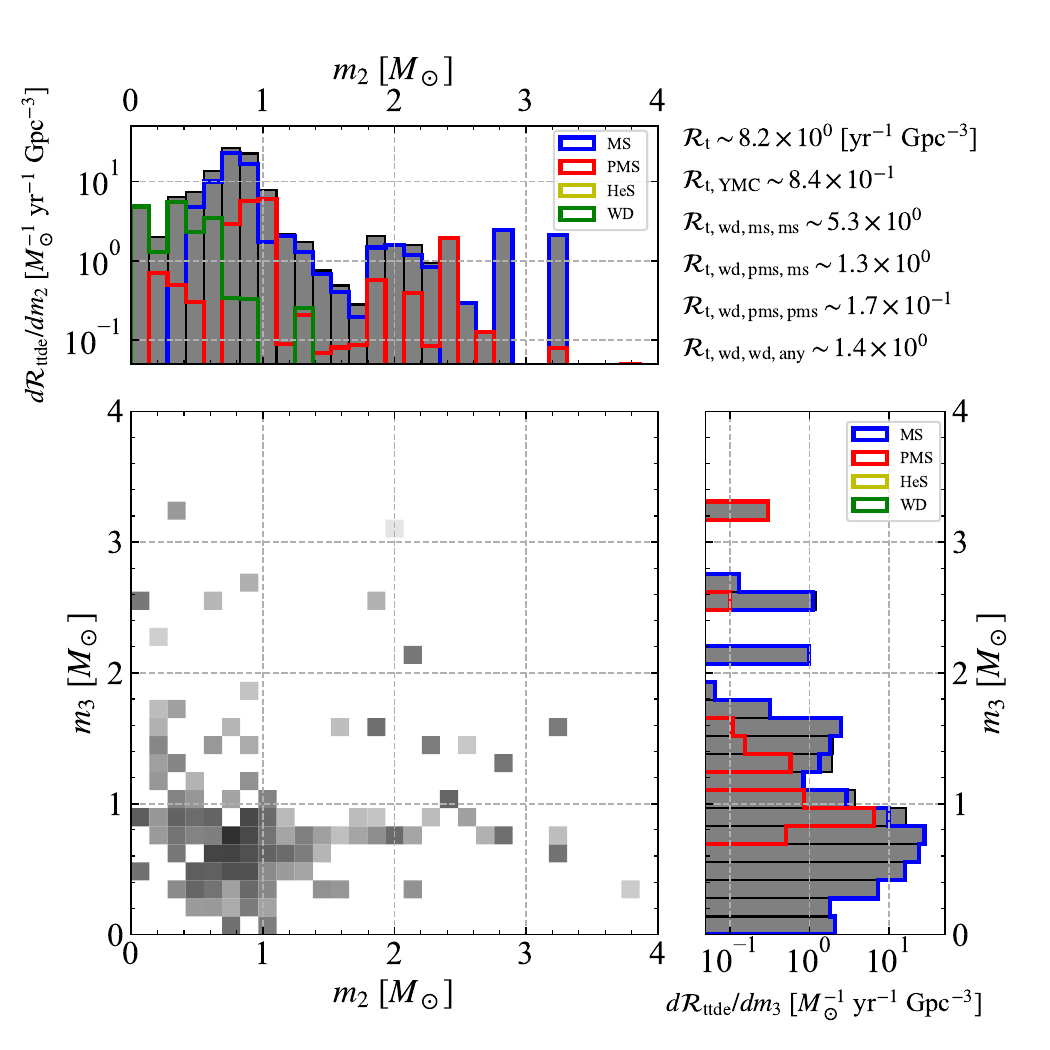}
  \caption{Triple TDE rate density in the current GCs and YMCs. We
    sort three disrupted stars in the order of ONe WD, CO WD, He WD,
    HeS, PMS, and MS, and in the descending order of masses if the
    stellar types are the same. We denote the first and second
    companion masses as $\mcp{2}$ and $\mcp{3}$, respectively. On the
    top-right corner, we show the rate densities of all triple TDEs
    ($\rtdet$), those in YMCs ($\rtdetk{YMC}$), those with WD and two
    MSs ($\rtdetk{wd,ms,ms}$), those with WD, MS, and PMS
    ($\rtdetk{wd,ms,pms}$), those with WD and two PMSs
    ($\rtdetk{wd,pms,pms}$), and those with two WDs and any type star
    ($\rtdetk{wd,wd,any}$). Bottom left: Relative abundance of the
    rate density per $\mcp{2}$ and per $\mcp{3}$. Top left: The rate
    density per $\mcp{2}$.  Gray-filled histograms indicate all the
    triple TDEs. Colored-open histograms indicate stellar types of
    companion stars, such that blue-, red-, yellow-, and green-open
    histograms are MSs, PMSs, HeSs, and WDs, respectively. Bottom
    right: The rate density per $\mcp{3}$. Color codes are the same as
    the top left panel.}
  \label{fig:tdeTriple}
\end{figure}

Figure \ref{fig:tdeTriple} shows the triple TDE rate density. The
total rate density is $\sim 8.2~\yrgpc$, similar to the double TDE
rate density, and only $\sim 1.7$ \% of the total WD TDE rate
density. The major part is dominated by WD TDEs with MS and PMS
TDEs. Interestingly, a WD TDE can accompany another WD TDE in triple
TDEs. $17$ \% of triple TDEs (or $0.35$ \% of the total WD TDEs) can
include double WD TDEs.

The WD mass distribution in the triple TDE rate density is similar to
that in the double TDE rate density (see the bottom right panel of
Figure \ref{fig:tdeDouble}). The mass distributions of non-WD stars in
triple TDEs are also similar to that in double TDEs. There are peaks
at $\sim 1.0 \msun$. The mass of $\sim 1.0 \msun$ corresponds to the
turn-off mass at $\sim 12$ Gyr.

\begin{figure*}
  \includegraphics[width=1.5\columnwidth]{\fdir/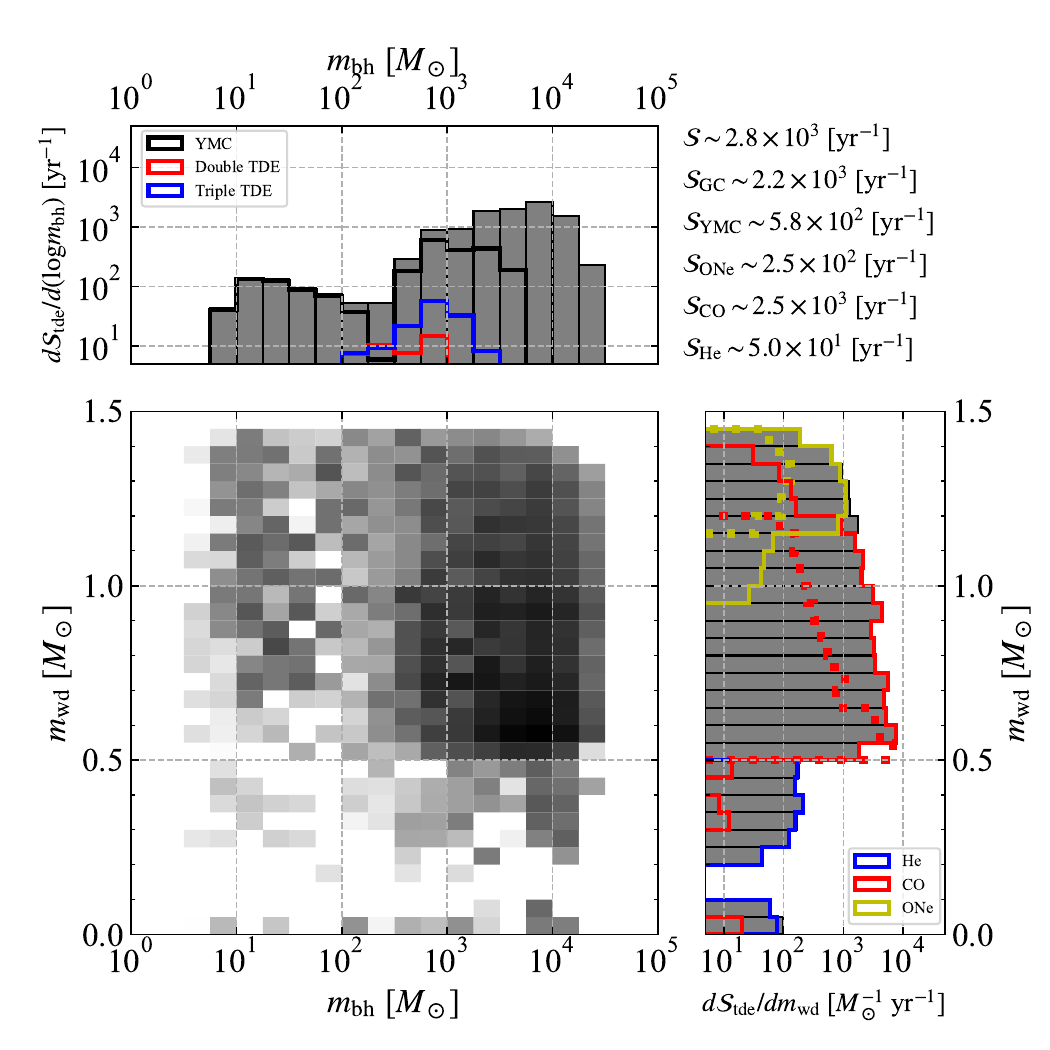}
  \caption{The same as Figure \ref{fig:tdeTotal} except that the WD
    TDE rate density in the local universe is replaced with the WD TDE
    rate within a redshift of $0.3$ (the volume-limited WD TDE rate).}
  \label{fig:tdeVolumeLimited}
\end{figure*}

Figure \ref{fig:tdeVolumeLimited} shows the WD TDE rate within a
redshift of $0.3$ (the volume-limited WD TDE rate), obtained from
Eq. (\ref{eq:VolumeLimitedRate}) for $z_{\max}=0.3$. The reason for
making this figure is that WD TDEs can be observed in the low-redshift
universe. We find small differences between the WD TDE rate density
and volume-limited WD TDE rate. The contribution of GCs in the
volume-limited WD TDE rate is smaller than in the WD TDE rate
density. YMCs are more actively formed in a higher-redshift
universe. Moreover, IMBHs in GCs do not yet end their growth in the
redshift of $0.3$. The relative abundance of CO WDs with $\sim 0.5
\msun$ decreases from the WD TDE rate density to the volume-limited WD
TDE rate. Such CO WDs are not yet formed, because the lifetimes of
their progenitors are longer than the universe age at the redshift of
$0.3$. Thus, more massive WDs become more dominant in the
volume-limited WD TDE rate.

\section{Comparison with previous studies}
\label{sec:Comparison}

In section \ref{sec:Results}, we see that the total WD TDE rate
density in GCs is $\sim 4.5 \times 10^2~\yrgpc$. On the other hand,
\cite{2020SSRv..216...39M} have estimated that the total rate density
is $\sim 4 \fbhgc~\yrgpc \sim 0.8~\yrgpc$, if GCs in dwarf galaxies
are included. Here, we adopt $\fbhgc = 0.19$ for the fraction of GCs
with IMBHs to the total GCs in the local universe as described in
section \ref{sec:Method}. Our current GC number density is $15.4~{\rm
  Mpc}^{-3}$. \cite{2020SSRv..216...39M} have supposed that the
current GC number density is $5.35~{\rm Mpc}^{-3}$. Although our
current GC number density is 3 times larger than theirs, the
difference between assumed GC number densities cannot explain the
difference between the derived rate densities. This means that we get
a WD TDE rate in each GC $\sim 100$ larger than adopted by
\cite{2020SSRv..216...39M}. They have adopted a WD TDE rate in each GC
calculated by \cite{2004ApJ...613.1143B}. In this section, we will
make clear the reason why our WD TDE rate in each GC is much larger
than obtained by \cite{2004ApJ...613.1143B}.

We can understand a WD TDE rate in each cluster, using the loss cone
theory developed by \cite{1976MNRAS.176..633F}. Moreover, we can also
interpret the difference between previous and our results in the
framework of the loss cone theory. Here, we first introduce the loss
cone theory briefly.

In the loss cone theory, a TDE occurs when a star keeps a loss-cone
orbit on its orbital time without diffused by two-body relaxation.
Based on the theory, \cite{2004ApJ...613.1143B} \citep[see
  also][]{2004ApJ...613.1133B} have derived their eq.~(7), the
description of a TDE rate in a cluster ($D$), as
\begin{align}
  D = \kd G^{1/2} \left[ \frac{\rstar^{9-4\alpha} \nu_0^7
      \mstar^{(16/3)\alpha-5}}{\mbh^{(19/3)\alpha-9}}
    \right]^{1/(8-2\alpha)}, \label{eq:TdeRateB04}
\end{align}
where $\kd$ is a dimensionless coefficient. We assume that a cluster
harbors an IMBH, and that the IMBH is surrounded by a stellar cusp in
which the stellar number density is written as $n = \nu_0
r^{-\alpha}$, where $n$ is the stellar number density, and $r$ is the
distance from the IMBH. If the IMBH mass is sufficiently smaller than
a cluster mass, the stellar cusp is contained in a constant density
core with density $\nc$ at the influence radius of the IMBH
$\rinfl$. The influence radius can be written as
\begin{align}
\rinfl = \frac{G\mbh}{\gamma\vc^2} \label{eq:InfluenceRadius}
\end{align}
where $\vc$ is the velocity dispersion in the core, and $\gamma$ is a
numerical factor. Finally, we can rewrite Eq.~(\ref{eq:TdeRateB04}) as
\begin{align}
  D = \kd G^{1/2+\alpha\beta} \gamma^{-\alpha\beta} \rstar^{2-\beta}
  \mstar^{(-8/3)+(7/3)\beta} \mbh^{(19/6)+(\alpha-7/3)\beta}
  \nc^{\beta} \vc^{-2\alpha\beta}, \label{eq:TdeRateT20}
\end{align}
where $\beta = 7/[2(4-\alpha)]$.

\begin{figure}
  \includegraphics[width=\columnwidth]{\fdir/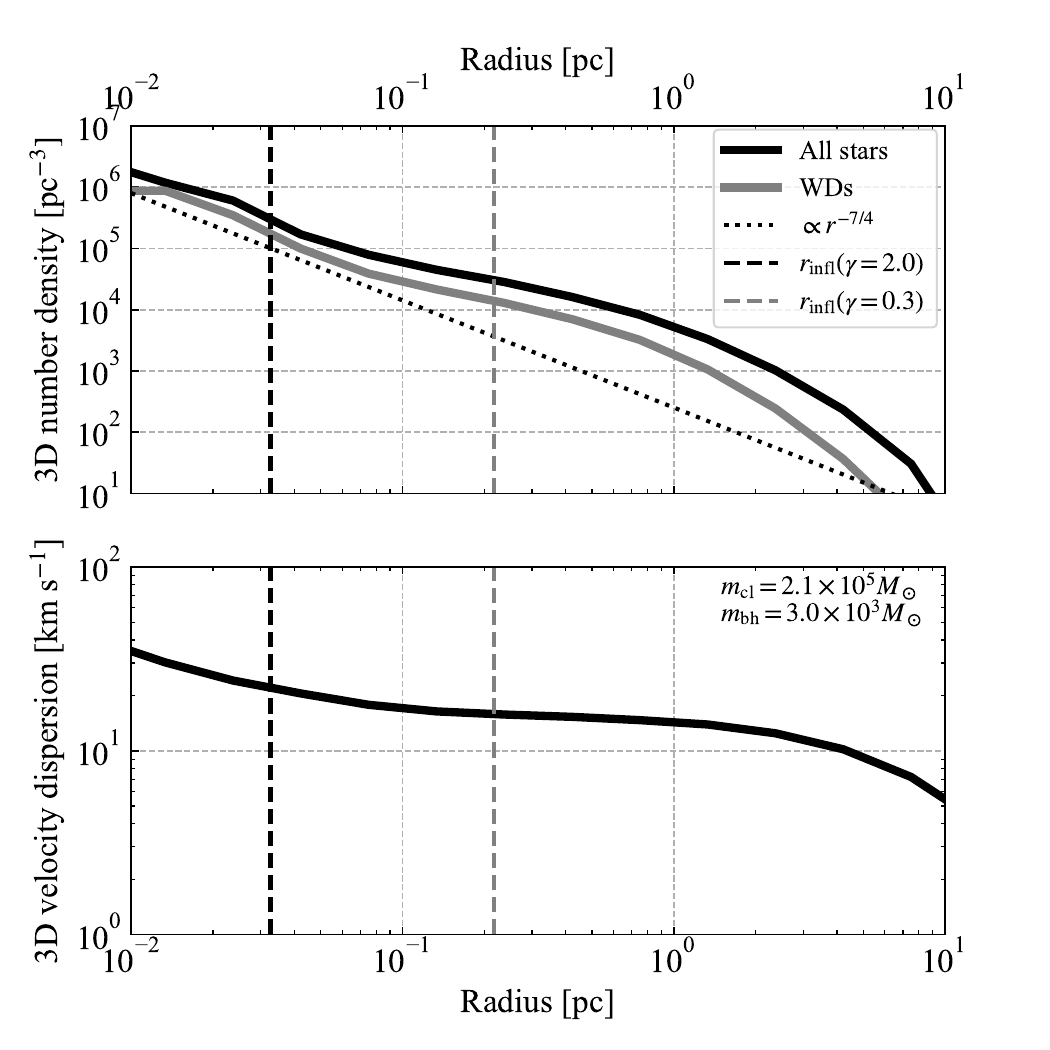}
  \caption{Profiles of 3D stellar number density (top) and 3D velocity
    dispersion (bottom). The age and initial conditions of the cluster
    is the same as in Figure \ref{fig:massWdAsComponent}. The current
    cluster mass and maximum BH mass are indicated in the bottom panel.
    Black- and gray-solid lines indicate profiles of all stars and
    WDs, respectively. The dotted line shows a stellar number density
    proportional to $r^{-7/4}$. The vertical black- and gray-dashed
    lines draw $\rinfl$ for $\gamma=2$ and $0.3$, respectively, where
    $\vc=15$~km~s$^{-1}$.}
  \label{fig:influenceRadius}
\end{figure}

We can adopt $\alpha=7/4$ for WDs in our clusters. This is consistent
with our 3D profile in Figure \ref{fig:influenceRadius}, and with the
results of \cite{2004ApJ...613.1143B}. This means that WDs form
so-called Bahcall-Wolf cusp \citep{1976ApJ...209..214B}. This is
because WD masses range from the minimum stellar mass ($\sim 0.1
\msun$) to the maximum stellar mass ($\sim 1.4 \msun$) except for
minor objects growing through stellar mergers. Note that a stellar
cusp made of MSs can have $\alpha \sim 3/2$ as seen in
\cite{2004ApJ...613.1143B}, since the maximum MS mass is at most $\sim
1.0 \msun$.

For $\alpha=7/4$, we can rewrite Eq.~(\ref{eq:TdeRateT20}) as
\begin{align}
  D &= \gamma^{-49/18} \kd G^{29/9} \rstar^{4/9} m^{26/27}
  \mbh^{61/27} \nc^{14/9} \vc^{-49/9} \label{eq:WdTdeRate1} \\
  &\sim 10^4 \; {\rm Gyr}^{-1} \; \left( \frac{\kd}{30} \right) \left(
  \frac{\gamma}{0.3} \right)^{-49/18} \left( \frac{\rstar}{1\rsun}
  \right)^{4/9} \left( \frac{m}{1\msun} \right)^{26/27} \nonumber \\
 &\times \left( \frac{\mbh}{10^3\msun} \right)^{61/27} \left(
  \frac{\nc}{10^5{\rm pc}^{-3}} \right)^{14/9} \left(
  \frac{\vc}{10{\rm km~s}^{-1}} \right)^{-49/9}. \label{eq:WdTdeRate2}
\end{align}
We adopt $\rstar=1\rsun$, $m=1\msun$, $\mbh=10^3\msun$,
$\nc=10^5$~pc$^{-3}$, and $\vc=10$~km~s$^{-1}$ for comparison with
\cite{2004ApJ...613.1143B}, despite that WD radii are typically $\ll 1
\rsun$. We choose $\kd=30$, which is consistent with the choice of
$\kd=65$ in \cite{2004ApJ...613.1143B}. We set $\gamma=0.3$ based on
stellar number density profiles. On the other hand,
\cite{2004ApJ...613.1143B} have set $\gamma=2$. As seen in Figure
\ref{fig:influenceRadius}, $\gamma$ is better for the choice. The
velocity dispersion starts increasing inside of the influence radius
with $\gamma=0.3$ (see the bottom panel). The influence radius with
$\gamma=2$ is completely inside of the stellar cusp. For confirmation,
we choose $\gamma=0.3$ (the vertical gray lines in Figure
\ref{fig:influenceRadius}), not $\gamma=2$ (the vertical black lines
in Figure \ref{fig:influenceRadius}). We just refer to $\gamma=2$ to
compare our results with previous studies. We will later show that
$\gamma$ critically affects the difference between the results of
\cite{2004ApJ...613.1143B} and ours.

We compare Eq.~(\ref{eq:WdTdeRate2}) with our simulation results. We
define a scaled TDE rate $\dscl$ as
\begin{align}
  \dscl &= D \left( \frac{\rstar}{1\rsun} \right)^{-4/9} \left(
  \frac{m}{1\msun} \right)^{-26/27} \left( \frac{\nc}{10^5{\rm
      pc}^{-3}} \right)^{-14/9} \left( \frac{\vc}{10{\rm km~s}^{-1}}
  \right)^{49/9}. \label{eq:WdTdeRateScaled}
\end{align}
We calculate $\dscl$ as follows. We calculate a rate of WD-BH
coalescence in each cluster during its age of $10$ -- $13$ Gyr. We
convert the rate into a WD TDE rate ($D$), using the corrective factor
described in the case I of Eq. (\ref{eq:CorrectionFactor}). We pick up
cluster profiles at their ages of $12$ Gyr, and obtain $\nc$ and $\vc$
as the stellar number density and velocity dispersion, respectively,
at a radius whose enclosed mass is the same as the maximum BH mass
aside from the BH itself. This assumption is valid only when the
stellar phase space distribution is isothermal, or only when the
radius is inside the stellar core. In fact, the radius is always
inside the stellar core for $\mbh \lesssim 10^4 \msun$.

\begin{figure}
  \includegraphics[width=\columnwidth]{\fdir/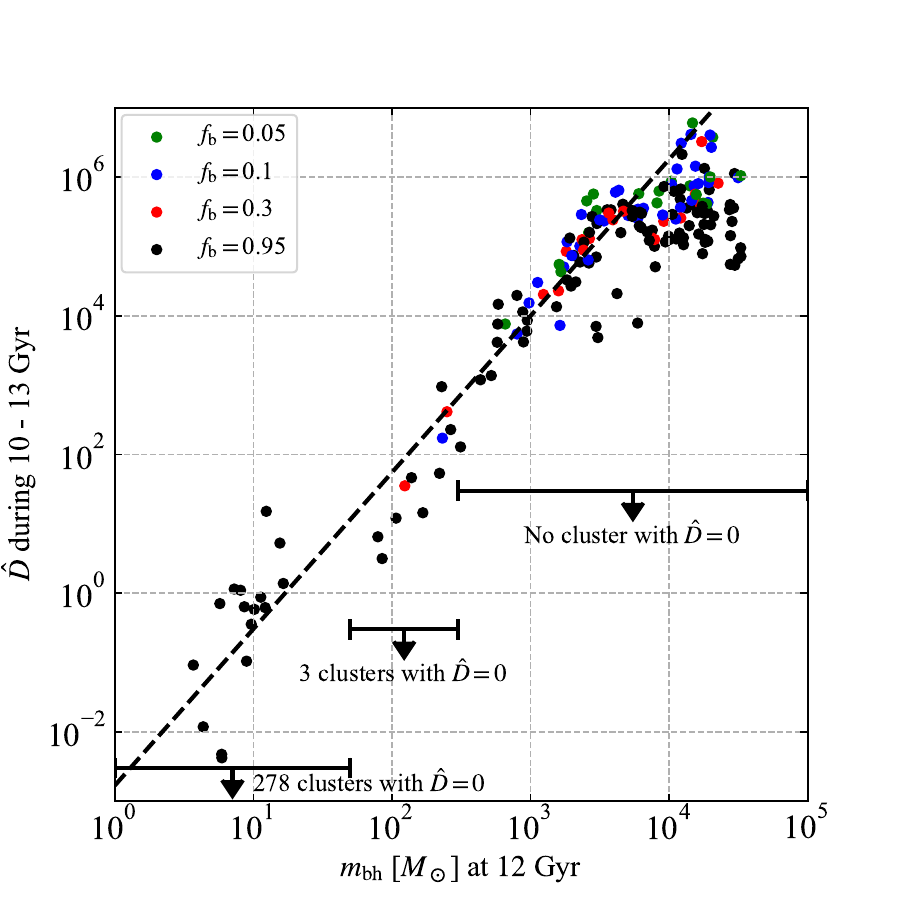}
  \caption{Scaled TDE rates during $10-13$~Gyr as a function of the
    maximum BH mass at $12$~Gyr. Clusters with $\mbh/\mcl>0.5$ are
    excluded. Color codes indicate the initial binary fraction of a
    cluster. The dashed line indicate
    Eq. (\ref{eq:WdTdeRateScaled}). There are many clusters without WD
    TDEs during $10-13$~Gyr ($\dscl=0$). The presence of these
    clusters are indicated by arrows with bars. The numbers of such
    clusters with $\mbh=1-50$ and $50-300$ $\msun$ are $287$ and $3$,
    respectively. There are no clusters with $\dscl=0$ and $\mbh>300
    \msun$.}
  \label{fig:rateScaled}
\end{figure}

We plot $\dscl$ in Eq. (\ref{eq:WdTdeRateScaled}) and $\dscl$ of each
cluster in Figure \ref{fig:rateScaled}, where we exclude clusters with
$\mbh/\mcl>0.5$, since we cannot define $\nc$ and $\vc$ for them. We
cannot plot clusters with $\dscl = 0$ due to the logarithmic
scale. Instead, the presence of such clusters is indicated by arrows
with bars. We find many clusters with $\dscl=0$ for $\mbh= 1-50
\msun$. On the other hand, there are only 3 clusters with $\dscl=0$
for $\mbh = 50-300 \msun$.  There are no clusters with $\dscl=0$ for
$\mbh > 300 \msun$.

We can see that $\dscl$ in Eq. (\ref{eq:WdTdeRateScaled}) agrees with
$\dscl$ of clusters for $10^2 \msun \lesssim \mbh \lesssim 10^4
\msun$. This means that WD TDE rates follow the loss cone theory for
this BH mass range. We expect that this is true for clusters with
$\mbh/\mcl>0.5$. The WD TDE rates should deviate from the theory when
$\mbh \lesssim 10^2 \msun$. Small BHs can wander in the core, since
they are easily perturbed by other stars. The theory cannot apply for
this case. In fact, there are many clusters with $\dscl=0$ as
indicated by arrows with bars in Figure \ref{fig:rateScaled}. It is a
coincidence that $\dscl$ in Eq. (\ref{eq:WdTdeRateScaled}) agrees with
$\dscl$ of clusters with non-zero $\dscl$ for $\mbh \lesssim 10^2
\msun$. However, this mass range has minor effects on the total WD TDE
rate density, since WD TDEs do not occur much for $\mbh \lesssim 10^2
\msun$. For $\mbh \gtrsim 10^4 \msun$, the scaled WD TDE rate appears
to be saturated. This is because the stellar cusps cover the stellar
cores in these clusters, and directly connect with the stellar halos,
not the stellar cores. Such saturation is also predicted by
\cite{2017MNRAS.467.4180S}. Eq. (\ref{eq:WdTdeRateScaled}) is also not
true for this case. Even for $\mbh \gtrsim 10^4 \msun$, several
clusters, in particular those with smaller $f_{\rm b}$, match better
with $\dscl$ in Eq. (\ref{eq:WdTdeRateScaled}). However, it would be a
coincidence. These clusters have high BH mass fractions, $\mbh/\mcl
\sim 0.4$. For such clusters, the whole clusters are stellar cusps
around the BHs. Eq. (\ref{eq:WdTdeRateScaled}) is not true for such
clusters.

Eq. (\ref{eq:WdTdeRate2}) may seem strange, because a $10^3 \msun$ BH
should be able to grow to a $10^5$ BH if it keeps accreting $1 \msun$
stars during $10$ Gyr. On the other hand, BHs in our cluster models
have at most $\sim 10^4 \msun$. This discrepancy may come from the
fact that BHs in our cluster models grow in the case II in
Eq. (\ref{eq:CorrectionFactor}), and Eq. (\ref{eq:WdTdeRate2}) is
deduced from the case I in Eq. (\ref{eq:CorrectionFactor}) (i.e. model
A). Actually, MS TDE rates calculated in the case I are larger than
those calculated in the case II by an order of magnitude (see
Figure \ref{fig:rateRawMs} in Appendix \ref{sec:OtherTDEs}). Thus, we
may underestimate BH masses in our cluster models by an order of
magnitude at most. However, the underestimate may not be so serious
for the following two reasons. First, BHs stop growing at $\sim 10^4
\msun$ due to depletion of surrounding stars as described
above. Second, we assume no mass loss in TDEs in our cluster models.
It should be stressed that the final fate of TDE debris is not fully
understood. On the one hand, some works indicate that up to half-mass
of the disrupted star could be lost in a TDE, and a considerable
fraction of such mass can be ejected from the BHs
\citep{2016MNRAS.461..948M, 2020MNRAS.492..686L}. On the other hand,
some recent simulations suggest that the fraction of ejected mass can
be rather limited \citep{2014ApJ...783...23G, 2015ApJ...804...85S}.

Although BH masses may be underestimated in our cluster models,
Eq. (\ref{eq:WdTdeRate2}) would be correct under the simulations where
such BHs are present in clusters. WD TDE rates match well with
Eq. (\ref{eq:WdTdeRate2}) as seen in Figure
\ref{fig:rateScaled}. Moreover, MS TDE rates are also in good
agreement with Eq. (\ref{eq:WdTdeRate2}) as seen in the left panel of
Figure \ref{fig:rateScaledMs} in Appendix \ref{sec:OtherTDEs}.

Interestingly, the scaled TDE rates do not much depend on the initial
binary fractions $f_{\rm b}$ for $\mbh = 10^2-10^4 \msun$. This means
that the initial binary fractions do not affect TDE rates if clusters
have $10^2-10^4 \msun$ BHs. On the other hand, all the clusters with
$\dscl > 0$ have $f_{\rm b}=0.95$ for $\mbh \lesssim 10 \msun$. These
clusters are more massive and more dense than other clusters for $\mbh
\lesssim 10 \msun$. Thus, they can have WD TDEs. The reason why such
high-mass and high-density clusters do not give rise to $\mbh \gtrsim
10^2 \msun$ during 12 Gyr is that they have high binary fractions
($f_{\rm b}=0.95$), and the binary interactions prevent rapid stellar
or BH mergers. Such clusters with low binary fractions ($f_{\rm b} \le
0.3$) already yield BHs with $\mbh \gtrsim 10^2 \msun$.

We can see that the $D$ in Eq. (\ref{eq:WdTdeRate2}) is $\sim 1000$
times larger than that in eq. (8) of
\cite{2004ApJ...613.1143B}. Before we discuss about the discrepancy
between their and our TDE rates ($D$), we point out inconsistency
between eqs. (7) and (8) of \cite{2004ApJ...613.1143B}. When we adopt
$\kd=65$ and $\alpha=1.55$ in the same way as
\cite{2004ApJ...613.1143B}, we obtain $D \sim 170~{\rm Gpc}^{-1}$, not
$D \sim 11~{\rm Gpc}^{-1}$. Actually, it is unclear which either
$\kd=65$ or eq. (8) is correct (in private communication with
H. Baumgardt). Since we find $\kd \sim 30$ from our results, we assume
that $\kd=65$ is correct. In this case, $D \sim 170~{\rm Gpc}^{-1}$ in
\cite{2004ApJ...613.1143B}. Nevertheless, our $D$ is still $\sim 100$
times larger than their $D$.

This discrepancy comes from $\gamma$ in
Eq.~(\ref{eq:InfluenceRadius}). \cite{2004ApJ...613.1143B} have chosen
$\gamma=2$ from their simulations (see their fig. 3), while we do
$\gamma=0.3$ from our simulation (see Figure
\ref{fig:influenceRadius}). If we substitute $\gamma=2$ into
Eq.~(\ref{eq:WdTdeRate2}), we get $D \sim 57$~Gyr$^{-1}$, which is
comparable to their $D$. This means that our clusters harbor $7$ times
larger stellar cusps than clusters of \cite{2004ApJ...613.1143B}, if
the two clusters have the same $\mbh$ and $\vc$. This results indicate
that our WD TDE rate agrees with a WD TDE rate of
\cite{2004ApJ...613.1143B} if we give the same BHs and stellar
cusps. In other words, the loss cone theory holds for both the cases.
However, if we give the same BHs and clusters, our WD TDE rate is
larger than theirs, since our cluster forms a larger stellar cusp than
their cluster. Reasons for this difference are unclear. Their and our
simulations contain a lot of different setup: IMFs, stellar evolution
models, cluster histories, numerical methods, and so on. We emphasize
that no BHs are pinned to the cluster centers in $N$-body simulations
of \cite{2004ApJ...613.1143B} and simulations of MOCCA-SURVEY Database
I. Thus in both the simulations BHs experience Brownian motion, which
can have great impacts on TDE rates \citep{1999MNRAS.309..447M}. As a
reference, we can obtain $\gamma \sim 0.9$ when we combine
Eq. (\ref{eq:InfluenceRadius}) and Eq. (A1) of
\cite{2018MNRAS.480.5060S} under the Bahcall-Wolf cusp, where the
latter equation is obtained from an approximate analytic
estimate. This value is intermediate between
\cite{2004ApJ...613.1143B} and our values. In either case, we need
detail comparisons among numerical and analytic models in order to
obtain the correct value of $\gamma$. We stop finding out the reasons
here.

If we prepare a cluster with the same $\mbh$, $\nc$, and $\vc$, our WD
TDE rate density in the current GCs is 1000 larger than obtained from
eq. (8) of \cite{2004ApJ...613.1143B} without correction about
inconsistency between eqs. (7) and (8). However, we can also derive
$\mbh$, $\nc$, and $\vc$ from the database. We find that many clusters
in the database have $\nc < 10^5~{\rm pc}^{-3}$ and $\vc > 10~{\rm
  km~s}^{-1}$ (see Figure \ref{fig:influenceRadius}, for
example). Thus, our TDE rate in the current GCs is only 100 larger
than the previous one.

Our WD TDE rate density in GCs ($\sim 4.5 \times 10^2~\yrgpc$) is much
larger than the total WD TDE rate density in GCs and dwarf galaxies
predicted by \cite{2016ApJ...819....3M} and
\cite{2020SSRv..216...39M}. They have estimated the total WD TDE
density as $\sim 10~\yrgpc$ or $\sim 300 \fbh~\yrgpc$, where $\fbh$ is
the BH occupation probability for dwarf galaxies. Here, we adopt $\fbh
\sim 0.03$, the lower limit of $\fbh$ derived by
\cite{2013ApJ...775..116R}, \cite{2014AJ....148..136M}, and
\cite{2018ApJ...868..152B}\footnote{There are large uncertainties in
$\fbh$; \cite{2015ApJ...799...98M} have suggested its lower limit is
$\sim 0.2$. Nevertheless, even if $\fbh=1$, our WD TDE rate density in
GCs is larger than the total WD TDE rate density in GCs and dwarf
galaxies.}. Their rate density is dominated by WD TDEs in dwarf
galaxies. They have assumed that WD TDEs in a dwarf galaxy with a BH
happen $1000$ times more efficiently than a GC with a BH. Note that
\cite{2020MNRAS.495.1061F} have pointed out that WD TDEs can be in
triple stellar systems, however its rate density is a fraction of the
WD TDE rate density in GCS and dwarf galaxies described above.  On the
other hand, we find that the WD TDE rate density in the current GCs is
100 larger than those obtained by them, and exceeds the total WD TDE
rate density in the current GCs and dwarf galaxies obtained by
them. This is because our WD TDE rate in each GC is comparable to
their WD TDE rate in each dwarf galaxy, and the current GC number
density is $10$ -- $100$ times larger than the current dwarf galaxy
number density. When we take into account our WD TDEs in YMCs, and
their WD TDEs in dwarf galaxies, we obtain $\sim 5 \times 10^2~\yrgpc$
for the WD TDE rate density in the current universe, and the rate
density is dominated by GCs. In summary, our WD TDE rate density in
the current universe is $\sim 50$ times larger than predicted
previously.

Finally, we remark that we may overestimate the WD TDE rate density,
since we overestimate BH growth for the following five reasons. First,
in a TDE, a BH should accrete a small fraction of a disrupted star
\citep{2015ApJ...804...85S, 2016MNRAS.461..948M, 2018ApJ...859L..20D,
  2019ApJ...882L..25L}. However, in star cluster models of the
database, a BH accretes the whole mass of a disrupted star. Thus, BHs
grow in the star cluster models much more than in reality. Second, we
do not account for GW recoil kicks induced by BH mergers. They can
eject BH merger remnants from the stellar cores and the clusters. They
can inhibit BH growth. As described in Appendix
\ref{sec:PessimisticCase} (model C), the WD TDE rate density decreases
by about 5 times when the GW recoil kicks are considered. Third, we
calculate the WD TDE rate density according to the case I of
Eq. (\ref{eq:CorrectionFactor}). On the other hand, the WD TDE rate
density decreases by about 5 times when we adopt the case II (see
model B in Appendix \ref{sec:Non-correctedWDTDERate}). Thus, we
overestimate the WD TDE rate by at most 5 times. Fourth, we set the
maximum stellar mass to $100 \msun$ at the initial time. If we adopt a
higher value (say $150 \msun$) for the maximum stellar mass, we get
more many BHs. In this case, BHs are more likely to be ejected through
few-body interactions among themselves. This also prevents BH
growth. As seen in Eq. (\ref{eq:WdTdeRate2}), a TDE rate increases
with BH mass increasing. If we overestimate BH mass in a cluster, we
also overestimate a WD TDE rate in a cluster. Fifth, we may assume
larger GC number density ($\ngc$) than in reality. Since WD TDEs
dominate all the WD TDEs in the local universe according to our
results, we may overestimate the WD TDE rate density by about $10$
times. Therefore, we should keep in mind that the estimated WD TDE
rate is the upper limit, i.e. $\lesssim 5 \times 10^2~\yrgpc$. When we
consider the WD TDE rate density in model C (see Appendix
\ref{sec:PessimisticCase}) as the lower limit, the WD TDE rate can be
$90$-$500~\yrgpc$.

\section{Detectability}
\label{sec:Detectability}

We obtain the current WD TDE rate density of $\lesssim 5 \times 10^2
\yrgpc$. Most of them is responsible for $10^3$ -- $10^4 \msun$ BHs in
GCs. Disrupted WDs consist of $2.2$ \% He WDs, $88$ \% CO WDs, and
$10$ \% ONe WDs. Based on these results, we discuss about
detectability of WD TDEs. We use the same observational parameters as
\cite{2016ApJ...819....3M} and \cite{2020SSRv..216...39M} unless
specified otherwise.

WD TDEs can launch luminous jets, which can be detected by wide-field
monitors, such as {\it Swift}'s Burst Alert Telescope (BAT). We assume
that such jets have a luminosity of $\sim 10^{48}~{\rm erg~s}^{-1}$
and a beaming factor of $\sim 1/50$. Then, the BAT can detect the jet
up to redshift $z \sim 1$. This means the detectable volume is $\sim
150~{\rm Gpc}^3$. The BAT can cover 20 \% of the sky. We put them
together, and obtain the detection rate of the BAT as $\lesssim
300~{\rm yr}^{-1}$. This may not be consistent with the detection rate
of gamma-ray bursts by the BAT, $\lesssim 100~{\rm
  yr}^{-1}$. Moreover, ultra-long gamma-ray bursts, one of possible EM
counterparts of WD TDEs \citep[e.g.][]{2016ApJ...833..110I}, occupy a
small fraction of all gamma-ray bursts. There may be two reasons for
this inconsistency. First, we may overestimate the WD TDE rate density
as summarized in the final part of section
\ref{sec:Comparison}. Second, WD TDEs generate luminous jets less
efficiently than expected. For example, the beaming factor of TDEs can
be $0.1$ \citep{2011Natur.476..425Z, 2012ApJ...753...77C}.  WD TDEs
may have to satisfy special conditions for jet launching, similarly to
MS TDEs \citep{2013A&A...552A...5V, 2013ApJ...763...84B,
  2017MNRAS.464.2481G}.

A part of WD TDEs can trigger thermonuclear explosion of disrupted
WDs, and can be observed similarly to SNe Ia. Hereafter, we call this
``thermonuclear transients''. The LSST will be the most powerful to
search for such thermonuclear transients in the near future. If a $0.6
\msun$ CO WD experiences such thermonuclear explosion, the LSST can
detect it within a volume of $13.3~{\rm Gpc}^3$. The LSST has a sky
coverage of $\sim 0.5$. We suppose that $1/6$ of WD TDEs can have
thermonuclear explosion. Then, the detection rate is $\lesssim
550~{\rm yr}^{-1}$.

This detection rate is based on a WD TDE with a $0.6 \msun$ CO WD
disrupted by a $10^3 \msun$ BH. However, luminosity of a thermonuclear
transient depends on WD mass \citep{2016ApJ...819....3M,
  2020ApJ...890L..26K}. Moreover, the success rate of thermonuclear
explosion should be sensitive to both BH and WD masses, and WD
compositions \citep[e.g.][]{2009ApJ...695..404R, 2018MNRAS.477.3449K,
  2018ApJ...865....3A}. As seen in Figure \ref{fig:tdeTotal}, WD mass
ranges from $0.5 \msun$ to $1.4 \msun$ in a nearly uniform way, and BH
mass can be $10^3$ -- $10^4 \msun$ with the same probability. We
should keep in mind that the detection rate contains uncertainties due
to different BH and WD masses.

Thermonuclear transients will be helpful to identify WD TDEs. We
discuss about how they are observed. There are two previous studies to
investigate emission properties of such
transients. \cite{2016ApJ...819....3M} have examined a TDE of a $0.6
\msun$ CO WD, while \cite{2020ApJ...890L..26K} have focused on a TDE
of a $0.2 \msun$ He WD. Both of them have suggested that such
transients should be rapid and faint transients in optical and
ultraviolet bands. Their evolution timescale is $\sim 10$ days, which
is shorter than SNe Ia ($\sim 20$ days). This is because their photon
diffusion timescale is shorter due to their smaller total mass ($\sim
0.1$ -- $0.5 \msun$) than SNe Ia ($\sim 1.0$ -- $1.5 \msun$). Their
absolute magnitude is $\lesssim -18$, which is fainter than SNe Ia
($\sim -19$ -- $-20$). They yield only $\lesssim 0.1 \msun$ of
radioactive nuclei, such as $^{56}$Ni, while SNe Ia have $^{56}$Ni of
$\sim 0.6 \msun$.

However, these results may not apply for a significant fraction of WD
TDEs. As seen in Figure \ref{fig:tdeTotal}, 20 \% of WD TDEs contain
$\gtrsim 1.0 \msun$ WDs. Such WD TDEs have long evolution timescale,
similar to SNe Ia. If they yield a large amount of $^{56}$Ni, even
their light curves can resemble those of SNe Ia. In this case, we need
not only photometric observations but also spectroscopic observations
in order to identify WD TDEs. WD TDEs can have Doppler shift of their
spectral lines up to $\sim 10,000~{\rm km~s^{-1}}$ due to their
orbital motion around BHs
\citep{2016ApJ...819....3M,2020ApJ...890L..26K}. Half of $\gtrsim 1.0
\msun$ WDs (or 10 \% of all WDs) are ONe WDs. It should be instructive
to investigate how ONe WD TDEs are observed.

WD TDEs can be GW transients. LISA can detect only WD TDEs in the
local group \citep{2009ApJ...695..404R, 2018ApJ...865....3A}, since
the LISA band is different from GW frequency of WD TDEs. Thus, the
detection rate is extremely low, $\lesssim 10^{-5}~{\rm yr}^{-1}$. On
the other hand, GW observatories sensitive to decihertz GWs, such as
DECIGO, can discover WD TDEs even in the high-redshift universe
\citep{2020CQGra..37u5011A}.

\cite{2017PhRvD..96f3007Z} have claimed that jets associated with ONe
WD TDEs can be sources of ultra high-energy cosmic rays. This model
requires a ONe WD TDE rate of $0.1$ and $10~\yrgpc$
\citep[][respectively]{2017PhRvD..96j3003A, 2018NatSR...810828B},
depending on the baryon loading of the jets. Our ONe WD TDE rate is
$\lesssim 50~\yrgpc$, large enough for this model. However, our CO WD
TDE rate density is $\sim 10$ times larger than our ONe WD TDE
rate. Thus, our WD TDE rate density may not be consistent with this
model, unless there are some processes to prevent CO WD TDEs from
yielding ultra high-energy cosmic rays.

Finally, we mention multiple TDEs with at least one WD TDE. The
fraction of such multiple TDEs to the total WD TDEs is only $\sim 2$
\%. However, they are quite interesting, since they have distinct
features owing to multiple emission components. Conversely, if an MS
TDE is found, it is worth examining the presence of a hidden WD
TDE. The discovery of such a hidden WD TDE will make the presence of
an IMBH conclusive.

\section{Summary}
\label{sec:Summary}

We have investigated MOCCA-SURVEY Database I, and obtain the local
rate density of WD TDEs in GCs and YMCs. The large data enables us to
calculate not only the total rate density, but also its
derivatives. We have found that the local rate density including WD
TDEs in dwarf galaxies is $\sim 90$-$500~\yrgpc$ in model A, 90 \% of
which are WD TDEs in GCs. This rate density is 9-50 times larger than
predicted by \cite{2016ApJ...819....3M} and
\cite{2020SSRv..216...39M}. WD TDEs in our cluster models happen
200-1000 times more efficiently than in cluster models of
\cite{2004ApJ...613.1143B}, used by \cite{2016ApJ...819....3M} and
\cite{2020SSRv..216...39M}. This is because our cluster models have 7
times larger stellar cusps than Baumgardt's models, even if their and
our cluster models have the same cluster cores. Our results may
increase the detectability of luminous jets, X-ray flares, and
thermonuclear transients induced by WD TDEs by 9-50 times, compared
with previous estimates. However, we have to make a caveat that we may
overestimate the WD TDE rate density because of inconsistency in the
mass growth of IMBHs as discussed in section \ref{sec:Comparison}.

We have also found that 20 \% of disrupted WDs have $\gtrsim 1.0
\msun$, and half of them are ONe WDs. If they experience thermonuclear
explosions, they can be observed as luminously and long as SNe
Ia. This is different from previous expectations that thermonuclear
transients should be more rapid and faint than SNe Ia because of
smaller ejecta masses and radioactive masses
\citep{2016ApJ...819....3M,2020ApJ...890L..26K}. Our results indicate
that we need to investigate observational features of massive WD
TDEs.

The rate density of ONe WD TDEs is $\lesssim 50~\yrgpc$, large enough
to yield observed ultra high-energy cosmic rays. However, CO WD TDEs
happens 8 times more frequently than ONe WD TDEs. If WD TDEs are the
origin of ultra high-energy cosmic rays, CO WD TDEs may need not to
create create ultra high-energy cosmic rays.

\section*{Acknowledgments}

We are very grateful to the anonymous referee for many constructive
suggestions. AT thanks A. Askar, H. Baumgardt, G. Fragione,
K. Kawana, M. MacLeod, and Y. Suwa for fruitful discussions. AT was
supported in part by Grants-in-Aid for Scientific Research (17H06360,
19K03907) from the Japan Society for the Promotion of Science. MG was
partially supported by the Polish National Science Center (NCN)
through the grant UMO-2016/23/B/ST9/02732. MAS acknowledges financial
support from the Alexander von Humboldt Foundation for the research
program ``The evolution of black holes from stellar to galactic
scales'', the Volkswagen Foundation Trilateral Partnership through
project No. I/97778 ``Dynamical Mechanisms of Accretion in Galactic
Nuclei'', and the Deutsche Forschungsgemeinschaft (DFG, German
Research Foundation) -- Project-ID 138713538 -- SFB 881 ``The Milky
Way System'', and funding from the European Union's Horizon 2020
research and innovation programme under the Marie Sk\l{}odowska-Curie
grant agreement No. 101025436 (project GRACE-BH, PI Manuel Arca
Sedda).

\section*{Data availability}

The data underlying this article can be obtained upon request to Mirek
Giersz (mig@camk.edu.pl) and after agreeing to the terms of the {\tt
  MOCCA} License. The license can be found in
\url{https://moccacode.net/license/}.


\appendix

\section{Other TDEs}
\label{sec:OtherTDEs}

\begin{figure*}
  \includegraphics[width=1.0\columnwidth]{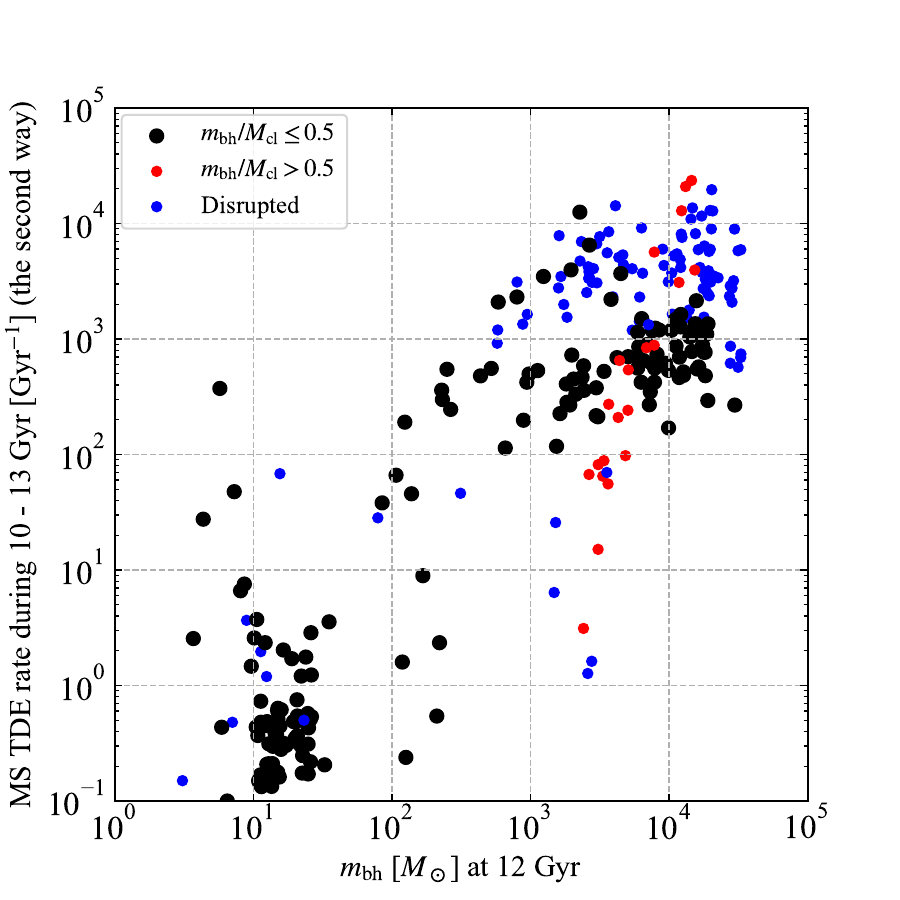}
  \includegraphics[width=1.0\columnwidth]{\fdir/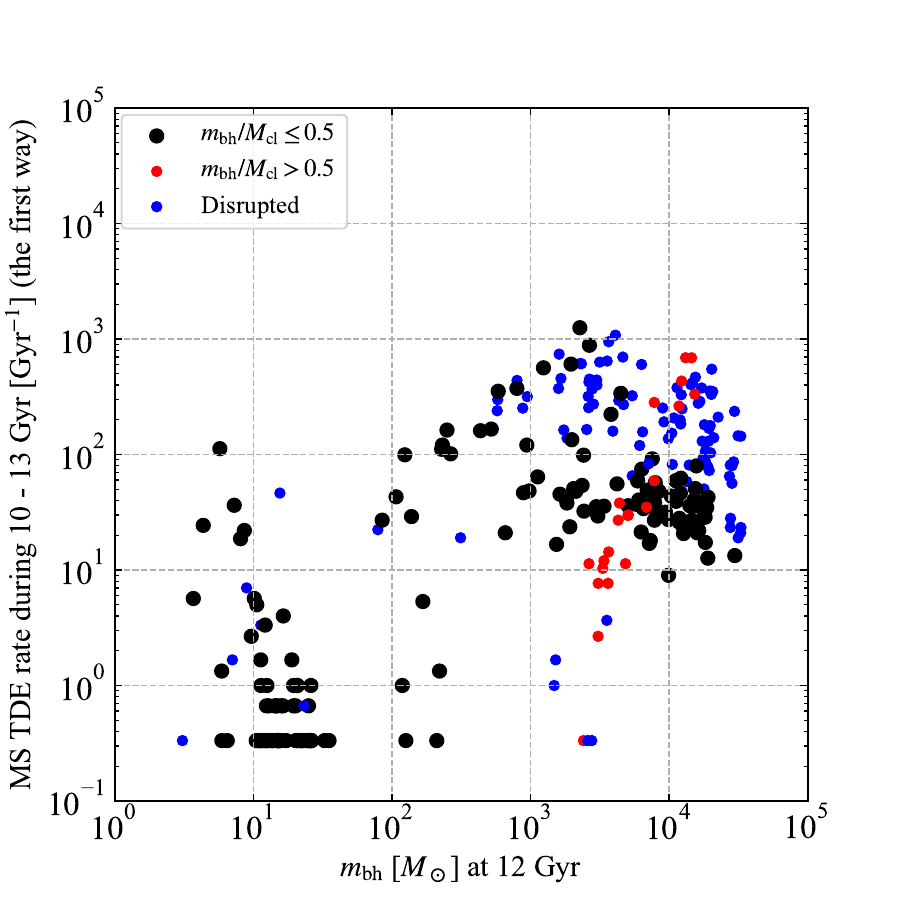}
  \caption{MS TDE rates during $10-13$~Gyr as a function of the
    maximum BH mass at $12$~Gyr in model A (left) and model B
    (right). Black and red circles indicate clusters with BH masses
    $\mbh/\mcl \le 0.5$ and $> 0.5$, respectively. Blue circles
    indicate clusters falling into galactic centers at $12$~Gyr, or
    clusters which are not disrupted at $10$~Gyr, but disrupted until
    $12$~Gyr.}
  \label{fig:rateRawMs}
\end{figure*}

In this section, we focus on other TDEs than WD TDEs, in particular MS
TDEs. Figure \ref{fig:rateRawMs} shows the MS TDE rates during
$10-13$~Gyr as a function of the maximum BH mass at $12$~Gyr. The left
and right panels indicate MS TDE rates calculated in model A and B in
Eq. (\ref{eq:CorrectionFactor}), respectively. MS TDE rates in model A
are larger than in model B by an order of magnitude for $\mbh \gtrsim
10^3 \msun$. This is because the tidal radii of MSs are larger than
the stellar radii of MSs.

\begin{figure*}
  \includegraphics[width=1.0\columnwidth]{\fdir/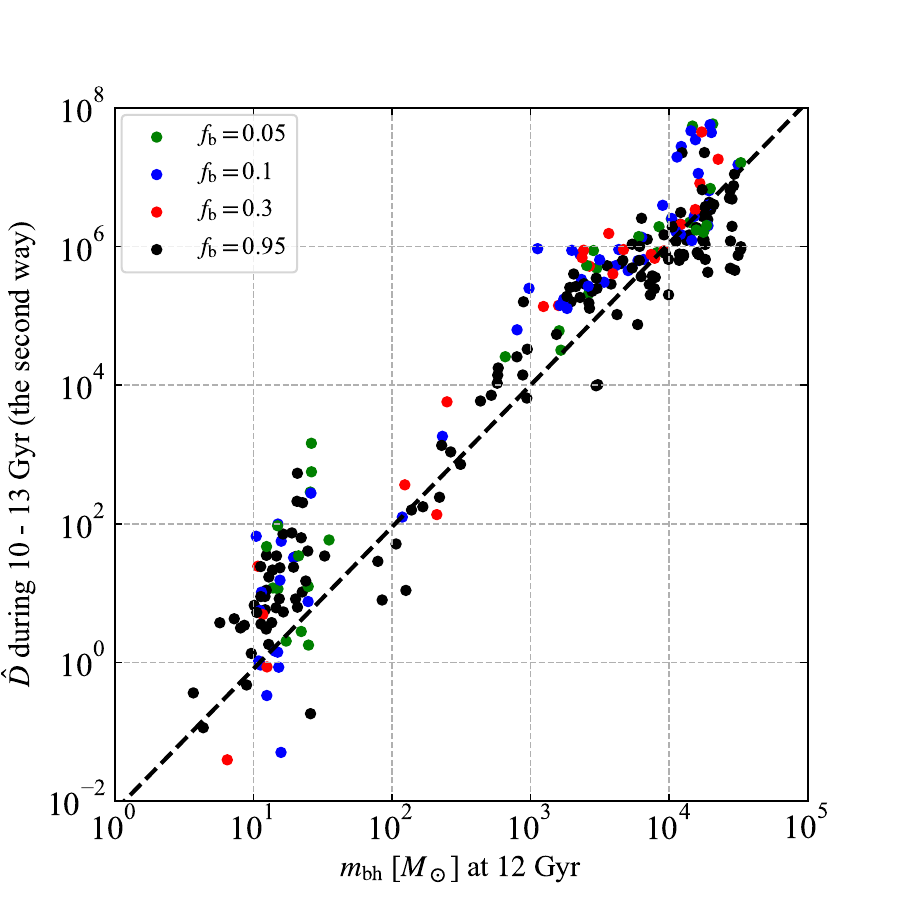}
  \includegraphics[width=1.0\columnwidth]{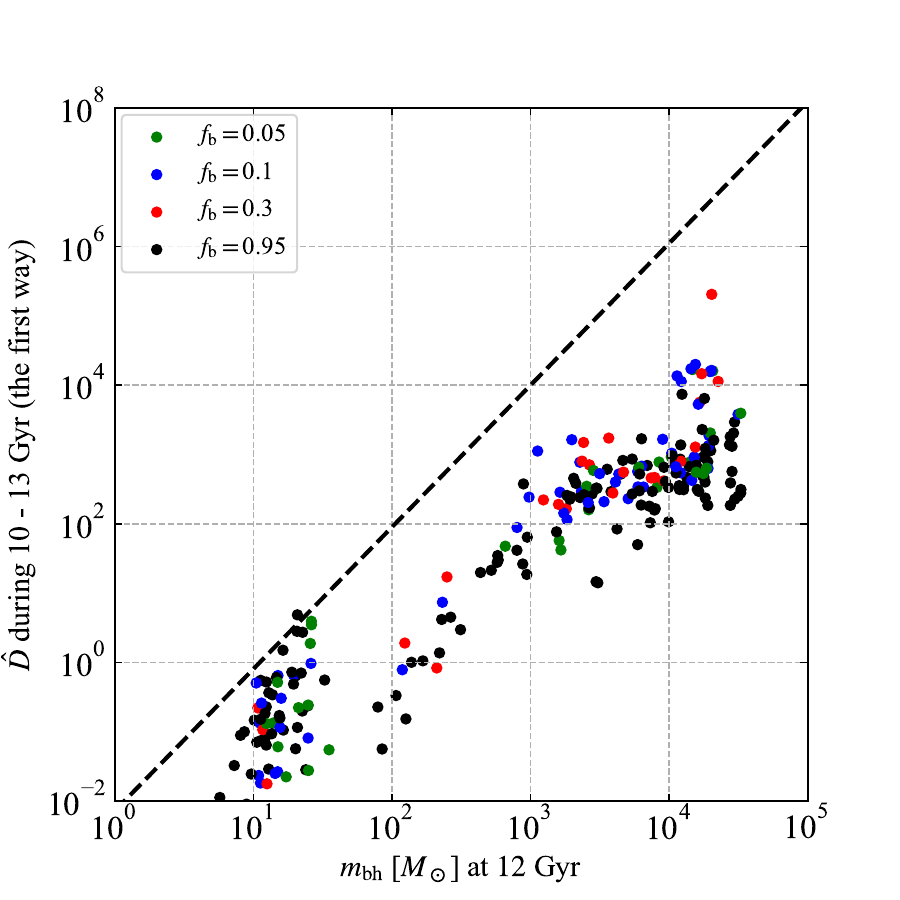}
  \caption{Scaled MS TDE rates during $10-13$~Gyr as a function of the
    maximum BH mass at $12$~Gyr in model A (left) and model B
    (right). Color codes and dashed lines are the same as Figures
    \ref{fig:rateScaled}.}
  \label{fig:rateScaledMs}
\end{figure*}

Figure \ref{fig:rateScaledMs} shows the scaled MS TDE rates during
$10-13$~Gyr as a function of the maximum BH mass at $12$~Gyr in model
A (left) and model B (right). The scaled MS TDE rates calculated in
model B does not fit to Eq. (\ref{eq:WdTdeRateScaled}). On the other
hand, the scaled MS TDE rates calculated in model A match well with
Eq. (\ref{eq:WdTdeRateScaled}), similarly to the scaled WD TDE rates
(see Figure \ref{fig:rateScaled}).

\begin{figure*}
  \includegraphics[width=1.0\columnwidth]{\fdir/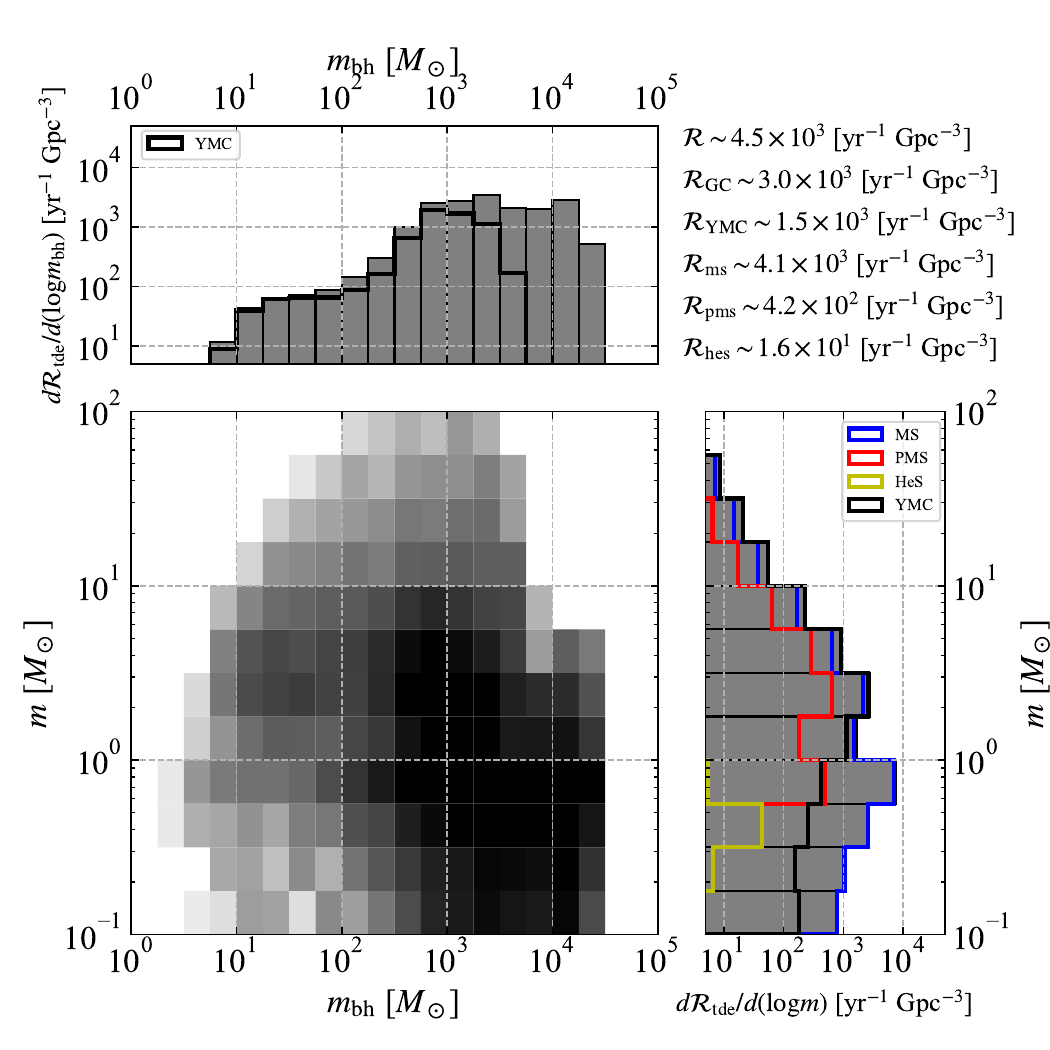}
  \includegraphics[width=1.0\columnwidth]{\fdir/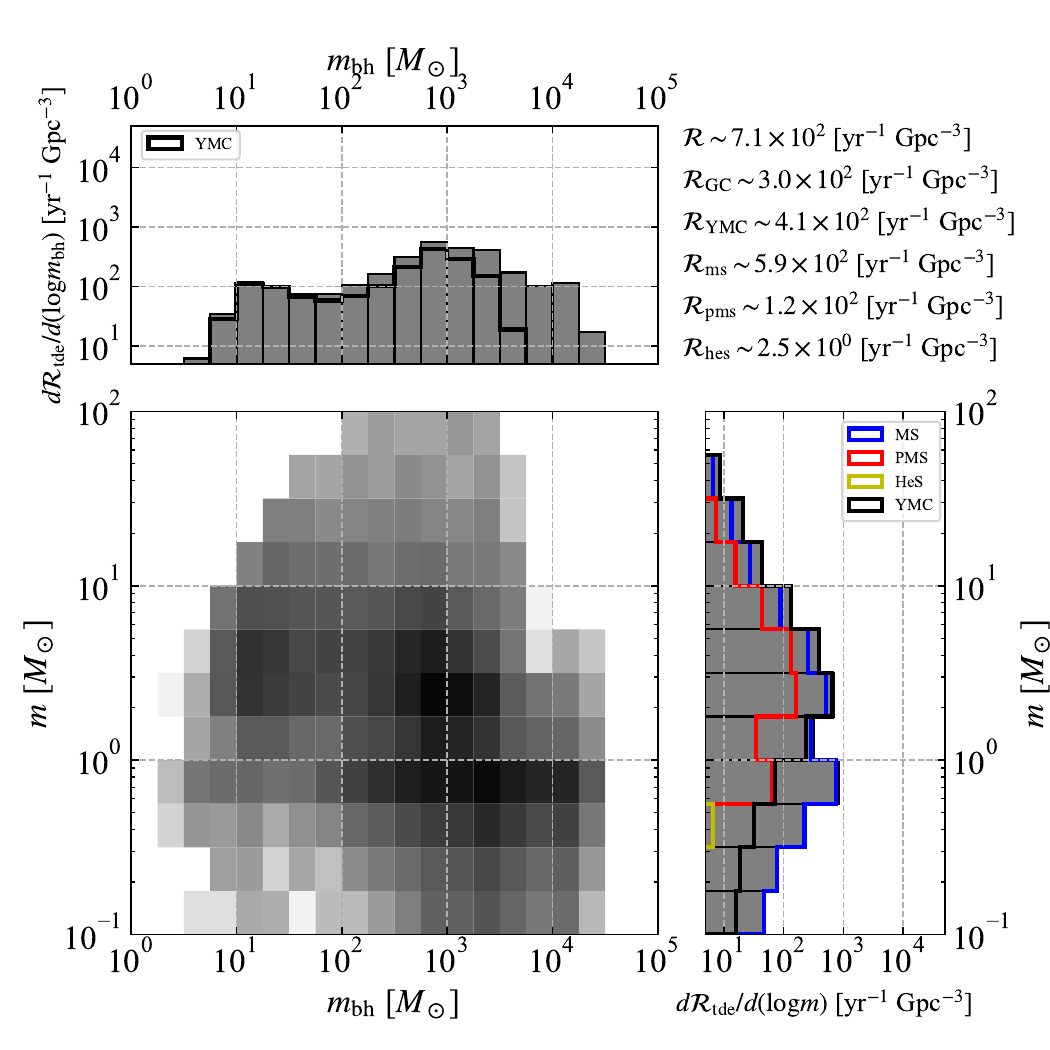}
  \caption{MS, PMS, and HeS TDE rate densities in the current GCs and
    YMCs calculated in model A (left) and model B (right). Since the
    left and right figures are the same except for the corrective
    factors, we only explain the left figure in the following. On the
    top-right conner, we show the rate densities of all MS, PMS, and
    HeS TDEs (${\cal R}$), these TDEs in GCs (${\cal R}_{\rm GC}$),
    theses TDEs in YMCs (${\cal R}_{\rm YMC}$), MS TDEs (${\cal
      R}_{\rm ms}$), PMS TDEs (${\cal R}_{\rm pms}$), and HeS TDEs
    (${\cal R}_{\rm hes}$).  Bottom left: Relative abundance of the
    rate density per dex stellar mass ($m$) per dex of $\mbh$ in a
    logarithmic color scale. Top left: The rate density per dex of
    $\mbh$. Gray-filled histograms indicate all the MS, PMS, and HeS
    TDEs, and black-open histograms these TDEs only in YMCs.  Bottom
    right: The rate density per dex stellar mass ($m$). Gray-filled
    and black-open histograms are the same as in the top left
    panel. Blue-, red-, and yellow-open histograms indicate MS, PMS,
    and HeS TDEs, respectively.}
  \label{fig:tdeTotalMs}
\end{figure*}

Figure \ref{fig:tdeTotalMs} shows the MS, PMS, and HeS TDE rate
densities calculated in model A (left) and model B (right). The MS TDE
rate density is much larger than other TDE rate densities including
the WD TDE rate density regardless of the choice of models A and
B. YMCs contribute to these TDEs more than GCs. Thus, the BH mass
distribution has a peak at $\sim 10^3 \msun$ in contrary to that of WD
TDEs. Note that BHs in YMCs can not grow up to $\sim 10^4 \msun$,
since YMCs have smaller mass than GCs at the initial time. YMCs
contain young massive stars, and thus the mass distribution of
disrupted stars has a peak at $2$ -- $3 \msun$. Owing to the presence
of such massive stars, these TDE rate densities in YMCs exceed those
in GCs in model B. The tendency described above is not sensitive to
models A and B.

\section{Non-corrected WD TDE rate}
\label{sec:Non-correctedWDTDERate}

\begin{figure*}
  \includegraphics[width=1.5\columnwidth]{\fdir/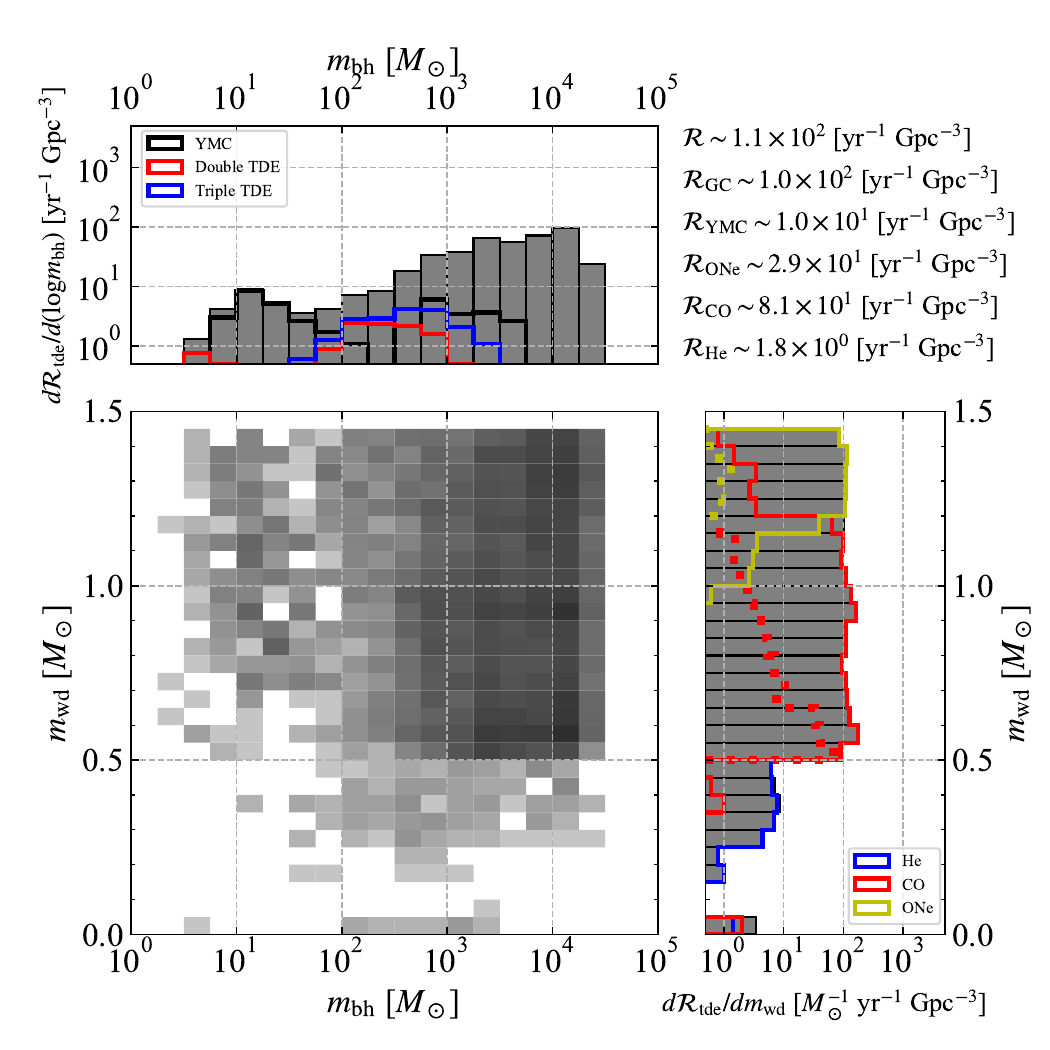}
  \caption{The same as Figure \ref{fig:tdeTotal}, except for model B.}
  \label{fig:tdeTotalRaw}
\end{figure*}

In this section, we show the WD TDE rate calculated in model B. Figure
\ref{fig:tdeTotalRaw} corresponds to Figure \ref{fig:tdeTotal}. The
total WD TDE rate density in model B is about 5 times smaller than in
model A. The corrective factor $\ctde$ in the case I of
Eq. (\ref{eq:CorrectionFactor}) is larger than unity, since
$\dtde/\dcoal$ is larger than unity as seen in Figure
\ref{fig:radiusCollTde}. Nevertheless, the total WD TDE rate density
is larger than estimated previously ($\sim 10~\yrgpc$) by an order of
magnitude. The BH mass distribution has a peak at a few $10^3 \msun$
in model A, while the peak shifts to $\sim 10^4 \msun$ in model
B. This is because the corrective factor $\ctde$ takes the maximum at
a few $10^3 \msun$ (see Figure \ref{fig:radiusCollTde}). Thus, our
conclusion that IMBHs dominate WD TDEs is robust. The WD mass
distribution in model B becomes more flat than in model A in the range
between $0.5$ and $1.4 \msun$. This reason is that $\dtde/\dcoal$
becomes larger for smaller WD mass. The contribution of ONe WDs is
significant even in model A, and is more pronounced in model B.

\begin{figure}
  \includegraphics[width=\columnwidth]{\fdir/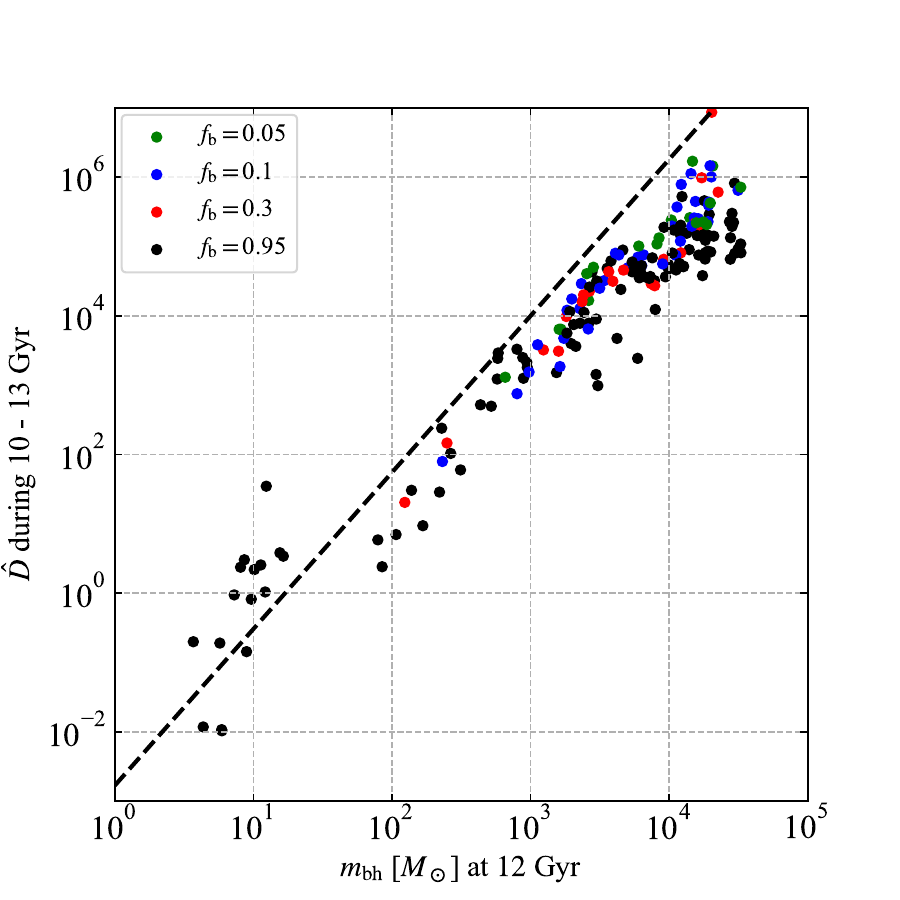}
  \caption{The same as Figure \ref{fig:rateScaled}, except for model
    B.}
  \label{fig:rateScaledRaw}
\end{figure}

Figure \ref{fig:rateScaledRaw} shows the scaled WD TDE rates in model
B, corresponding to Figure \ref{fig:rateScaled} in model A. Because
$\ctde$ is larger than unity in the case I of
Eq. (\ref{eq:CorrectionFactor}), the scaled WD TDE rates in model B
are smaller than in model A. Thus, the rates deviate from
Eq. (\ref{eq:WdTdeRateScaled}) downward. Nevertheless, the deviation
is small, a factor of about $5$, and the dependence on $\mbh$ is
similar to Eq. (\ref{eq:WdTdeRateScaled}) over several orders of
magnitude of $\mbh$. The theory deducing
Eq. (\ref{eq:WdTdeRateScaled}) should be correct, and
Eq. (\ref{eq:WdTdeRateScaled}) would be still helpful to estimate WD
TDE rates in star clusters.

\section{Pessimistic case}
\label{sec:PessimisticCase}

Here, we show WD TDEs in model C. In this model, we consider the
pessimistic case for IMBH formation. Before we explain the pessimistic
case, we describe the IMBH formation mechanisms.

\cite{2015MNRAS.454.3150G} have found that IMBHs can grow in either of
two ways, the so-called fast and slow scenarios. In the fast scenario,
IMBHs appear through mergers of stellar-mass BHs and massive stars,
where such massive stars are formed through runaway mergers. Since
this scenario needs massive stars, it happens only in a young phase of
a cluster. In the slow scenario, a $\sim 100 \msun$ BH is built up by
mergers of stellar-mass BHs and NSs, and grows through coalescences
with low-mass stars. Since the building up happens after gravothermal
core collapse, the IMBH appears only after a cluster age of about 1
Gyr.

Cluster simulations in MOCCA SURVEY Database I may overestimate and
overproduce IMBHs for different reasons between the fast and slow
scenarios. For the fast scenario, the cluster simulations assume that
stellar-mass BHs perfectly accrete massive stars through their
mergers. However, this assumption may overestimate IMBH masses. For
the slow scenario, the cluster simulations do not consider GW recoil
kicks after BH mergers. Thus, the building up may overproduce
IMBHs. In fact, a part of IMBHs fail to grow due to ejections of BHs
and NSs through GW recoil kicks \citep{2018MNRAS.481.2168M}.

In the pessimistic case, we divide IMBHs into those formed through the
fast and slow scenarios according to the time when BHs grow to $150
\msun$. If the time is less than $300$ Myr, IMBHs grow through the
fast scenario, and otherwise through the slow scenario
\citep{2019arXiv190500902A}. We exclude all the WD TDEs driven by
IMBHs formed through the fast scenario. This is because the accretion
efficiency of massive stars onto stellar-mass BHs may be quite
small. We ignore the presence of IMBHs with the probability of $60$,
$70$, and $90$ \% for clusters with $R_{\rm t}/R_{\rm h} = 50, 25$ and
$1$ (filling), respectively (see Table \ref{tab:MoccaSurveyDatabaseI})
according to the overall retention fractions shown in table 2 of
\cite{2018MNRAS.481.2168M}.

We generate 21 realizations, because IMBHs formed through the slow
scenario survive inside clusters in a stochastic way. Among the 21
realizations, the minimum, median, and maximum values of WD TDE rate
densities are $2.3 \times 10^1$, $8.9 \times 10^1$, and $1.8 \times
10^2$ $\yrgpc$, respectively. For $\fbhgc$, the minimum, median, and
maximum values are $0.011$, $0.019$, and $0.026$, respectively. The WD
TDE rate densities are weakly correlated with $\fbhgc$. The WD TDE
rate density in model C is typically less than in model A only by
about 5 times, despite that $\fbhgc$ decreases by about $10$
times. This is because underfilling clusters (those with $R_{\rm
  t}/R_{\rm h}=25$ and $50$) are easier to retain IMBHs and generate
WD TDEs.

\begin{figure*}
  \includegraphics[width=1.5\columnwidth]{\fdir/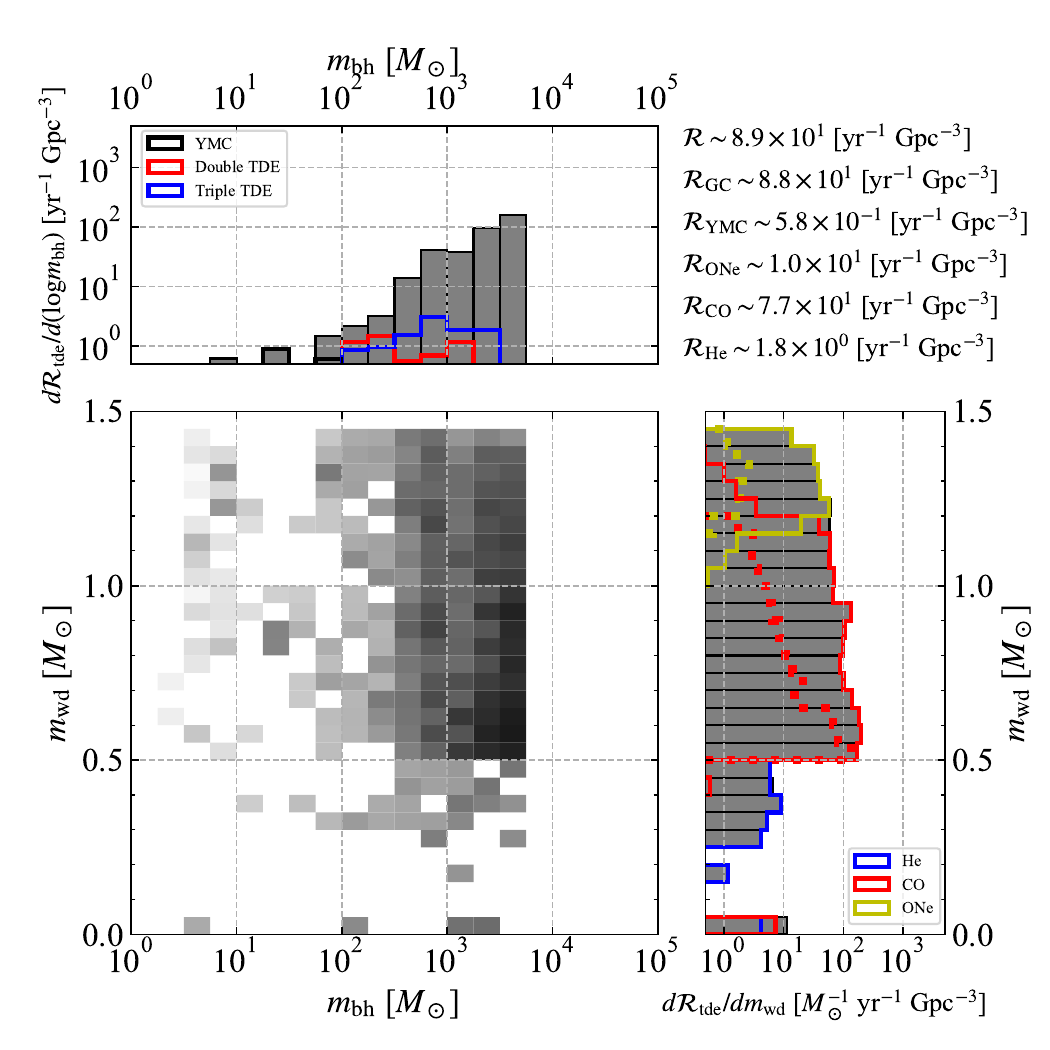}
  \caption{The same as Figure \ref{fig:tdeTotal}, except for a
    realization in model C. The realization has the median value of WD
    TDE rate densities among the 21 realizations.}
  \label{fig:tdeTotalPes}
\end{figure*}

Figure \ref{fig:tdeTotalPes} shows the WD TDE rate density and its
breakdowns for the realization with the median WD TDE rate density
among the 21 realizations. The striking feature is the absence of
IMBHs with $\sim 10^4$ $\msun$. Since such IMBHs are formed through
the fast scenario, they cannot contribute to the WD TDE rate density
in model C. Other IMBHs can survive inside clusters with a probability
described above. Thus, the WD TDE rate density also decreases in IMBHs
with $< 10^4$ $\msun$. The shape of the WD mass distribution is
similar to that in the optimistic case.

\bsp	
\label{lastpage}
\end{document}